\documentclass[epj, nopacs]{svjour}
\usepackage[T1]{fontenc}
\usepackage{cite}
\usepackage{hyperref}
\usepackage{blindtext}
\usepackage{physics}
\usepackage{empheq}
\usepackage{amsmath}
\usepackage{amssymb}
\usepackage{bigints}
\newcommand{\bibcommenthead}[1]{}

\begin{document}

\title{\emph{n}-dimensional non-commutative GUP quantization and application to the Bianchi I model}
\titlerunning{\emph{n}-dimensional non-commutative GUP quantization and application to the Bianchi I model}

 \author{Sebastiano Segreto\inst{1,3} \thanks{sebastiano.segreto@uniroma1.it} \and Giovanni Montani\inst{1,2}\thanks{giovanni.montani@enea.it}}
 \authorrunning{S. Segreto \and G. Montani}

 \institute{Physics Department, 'Sapienza' University of Rome, P.le Aldo Moro 5, Rome, 00185, Italy, \and ENEA, FSN-FUSPHY-TSM R.C. Frascati, Via E. Fermi 45, Frascati, 00044, Italy \and INFN, Sezione di Roma 1, P.le A. Moro 2, Rome, 00185, Italy }

\abstract{We analyse a $n$-dimensional Generalized Uncertainty Principle (GUP) quantization framework, 
    characterized by a non-commutative nature of the configurational variables.
    First, we identify a set of states which are maximally localized only along a single direction, at the expense of being less localized in all the other ones.
    Subsequently, in order to recover information about localization on the whole configuration space, we use the only state of the theory which exhibits maximal localization simultaneously in every direction to construct a satisfactory quasi-position representation, by virtue of a suitable translational operator.
    The resultant quantum framework is then applied to model the dynamics of the Bianchi I cosmology. The corresponding Wheeler-DeWitt equation is reduced to Schr\"odinger dynamics for the two anisotropy degrees of freedom, using a WKB representation for the volume-like variable of the Universe, in accordance with the Vilenkin scenario.
    The main result of our cosmological implementation of the constructed quantum theory 
    demonstrates how the dynamics of a wave packet peaked at some point in the configuration space represented in the quasi-position variables favours as the most probable configuration exactly the initial one for a relatively long time, if compared with the ordinary quantum theory.
    This preference arises from the distinct behavioral dynamics exhibited by wave packets in the two quantum theories.}

\maketitle

 \section{Introduction}

    As widely acknowledged, any quantum theory of gravity lays down its foundations in the attempt to reconcile General Relativity with Quantum Mechanics. Despite the lack of a comprehensive theory and the absence of consensus regarding the predominant approach, the idea that addressing this issue demands a profound reconsideration of the fundamental structure of space-time is broadly spread and accepted.
    Consequently, several models and theories based on a space-time structure fundamentally different with respect to the one arising in General Relativity have been constructed so far \cite{Hossenfelder:2012jw}.
    While certain frameworks aim to construct entirely new foundational paradigms, there exist different formulations that can be considered effective models.
    These models are capable of reproducing and making manifest some essential aspects derived from a potentially unspecified ultimate theory that, in principle, accounts for the altered structure of space-time.

    Generalized uncertainty principle (GUP) theories are quantum non-relativistic theories based on a deformation of the ordinary Heisenberg's uncertainty principle (HUP) and they do belong to this last category.

    There exist several arguments, stemming from String and Gedanken experiments \cite{Maggiore:1993rv}, which suggest that a modification of the HUP is indeed necessary at some level.

    The first examples in which the subject has been comprehensively investigated and analysed can be found in \cite{Kempf:1993bq, Kempf:1994su}, where considerations are specifically grounded in the low energy limits of string dynamics \cite{Amati:1988tn, Konishi:1989wk}.

    From these works clearly emerges that the most rigorous way to proceed in this direction is to deform the usual Heisenberg algebra between the quantum conjugate operators, hence modifying the canonical structure of ordinary quantum mechanics.
    According to the specific algebra, this operation results in two noteworthy consequences: firstly, the possible appearance of a nonzero absolute minimal uncertainty in the coordinate operators, which equips the theory with some kind of "minimal structure"; secondly, the possible emergence of a non-commutative "geometry", manifesting itself in the non-commutativity of the operators associated to the configurational variables.

    Despite it can help to gain useful insights, when this scheme is applied to the quantum dynamics of a particle, not surprisingly, the corresponding observables turn out to have negligible corrections. 
    This is essentially due to the fact that the scale at which the effect generated by the deformations of the algebra becomes relevant is usually Planckian.

    Nonetheless, the implementation of such generalized approaches in the Mini-superspace \cite{cianfrani2014canonical} of cosmological models \cite{Barca:2021epy, Barca:2021qdn} is potentially able to offer an interesting perspective to introduce cut-off physics effects on the Universe dynamics during the Planck era of its evolution. 

    Driven by these motivations, significant interest has emerged in the literature (see in particular \cite{Maggiore:1993kv, Detournay:2002fq} or more recently \cite{Segreto:2022clx, Bosso:2021koi, Bosso:2023aht} for general aspects as well as \cite{Pachol:2023bkv, Ali:2024tbd, Wojnar:2023bvv} for applications and relation with different formalisms) for the Generalized Uncertainty Principle (GUP) physics, in particular toward possible extensions of the original paradigm proposed in \cite{Kempf:1994su}.

    Among these, a formulation that possesses a certain degree of generality and hence significance is the one proposed in \cite{Maggiore:1993kv, Maggiore:1993zu}.
    More recently in \cite{Fadel:2021hnx}, it has been argued that this GUP theory, distinguished for the presence of a square root in the position-momentum commutator, is a viable extension of the theory, still preserving the existence of a non-zero minimal uncertainty for the coordinate operator. This statement has been amended in \cite{Segreto:2022clx}, where it has been demonstrated the necessity to deal with a compact momentum space in order to recover the emergence of a minimal structure in this deformed algebra.

    The effort to generalize these modified algebra theories also involved constructing consistent multi-dimensional extensions.
    In this context, the key consideration is that, as previously stated, modified Heisenberg algebras in the $n$-dimensional scenario can determine a non-commutativity in the coordinate operators.
    While there are studies exploring $n$-dimensional commutative GUP theories \cite{Kempf:1996nk}, there is an evident absence of analysis regarding their non-commutative counterparts.

    it is evident that, in a mini-superspace formulation of the GUP framework, the possibility for a higher-dimensional (and, in particular for a two-dimensional) case gains in relevance for the simple fact that the Universe dynamics can be in general described by three scale factors \cite{Montani:2009hju}.
    Specifically, referring to the Bianchi models class, these degrees of freedom are essentially a volume-like variable and two physical anisotropy variables \cite{Misner:1973prb}. 

    The novelty of the present study is the construction of a quantum non-commutative GUP theory in the case of a generic number of space coordinates, able to offer a consistent dynamical scenario in which it is possible to investigate the dynamics of the Bianchi I Universe in a WKB reduction \cite{Landau:1991wop} of its quantum dynamics à la Vilenkin \cite{Vilenkin:1988yd} (see also \cite{Kiefer.44.1067, DiGioia:2019nti, Maniccia:2021skz, Maniccia:2023cgv}). 

    As a first and fundamental step of our analysis, we define a family of states, within the $n$-dimensional GUP framework, able to reach the maximal localization with respect to one generic coordinate, while the remaining one unavoidably exhibits greater uncertainties.
    The states we construct from this procedure have a general structure, wherein the minimal uncertainty can apply to any of the $n$ coordinates. However, this consistently results in an anisotropic effect, leaving the other coordinates non-uniformly localized. This inherent anisotropy is a natural manifestation of the non-commutativity of our GUP theory.
    While these derived states prove useful in determining localization along a single direction, they fall short in mapping the entire configuration space. Consequently, to achieve a suitable generalization of the "quasi-position'' representation of the quantum particle dynamics \cite{Kempf:1994su, Bosso:2020aqm, Bosso:2021xsn}, we leverage the only isotropic with respect to the maximal localization properties state allowed within our proposed scenario.
    This state describes a particle located at the origin of the coordinate system, having simultaneously the same (minimal) uncertainty in all the space directions.

    By employing a properly defined translational operator, we can shift the expectation position value of this isotropic state to a generic point in the configuration space, finally obtaining a family of states which can serve as a generalization of the quasi-position basis. This way, we are able to properly describe the localization properties and the evolution of a wave packet in the quasi-position space.
    This method permits us to extract some phenomenological considerations and insights from our study when applied to the behaviour of the Bianchi I model. 

    We describe the Bianchi I cosmology via standard Misner variables \cite{Misner:1967uu}, as mentioned above.
    In order to deal with a Schr\"odinger equation for the quantum dynamics - which is, in principle, fixed by a Wheeler-DeWitt formulation \cite{Giovannetti:2021bqh, Giovannetti:2022qje} - we adopt a WKB scenario à la Vilenkin, in which the isotropic coordinate approaches a quasi-classical limit, while the two anisotropy coordinates explore a small quantum phase-space (for previous applications of the same type see 
    \cite{Battisti:2009qd, Chiovoloni:2020bmh, DeAngelis:2020wjp}). 
    Our formulation of the two-dimensional GUP paradigm is then implemented to describe the two anisotropy degrees of freedom as non-commutative coordinates and, since the quantization has to refer to the momentum representation,  we adopt the construction proposed in \cite{Giovannetti:2023hdd}. 
    The most relevant characteristic of our cosmological model lies in the tendency of dynamics to favour the initial configurations as the most probable states of our Universe, at least for a certain time. Specifically, the initial states are probabilistically favoured for longer time intervals compared to what happens in the ordinary theory. The greater significance of this fact becomes clear when considering an isotropic state as the initial condition.
    This behaviour is essentially due to the fact that the peak of our wave packets is moving "slowly" in time, exhibiting a certain degree of "inertia" in leaving the initial point of the quasi-anisotropies plane in which we have placed it, i.e. the quasi-position plane in the case of the anisotropy variables.
    It is clear that this signals a departure from ordinary quantum mechanics, highlighting how non-commutative GUP algebras can introduce novel and relevant effects in the dynamics of the early Universe and provide insights - perhaps in more comprehensive models - on some of the open problems in quantum cosmology.

    The paper is organized as follows: in Section \ref{sec_II} we introduce the $n$-dimensional non-commutative GUP Heisenberg algebra and we outline a functional analysis of the framework.
    In Section \ref{sec_III} we face and address the problem of the maximally localized states and their definition in a non-commutative setting. In Section \ref{sec_IV}, exploiting the results of the previous section, we proceed to define and construct a suitable generalization of the quasi-position basis and its relation with the momentum representation.
    In Section \ref{sec_V}, using Vilenkin's scheme for quantum cosmology, we apply the developed GUP quantum theory to the dynamics of the Bianchi I model and explore its behaviour, comparing it to what happens in the ordinary quantum scenario.
    Finally, in Section \ref{sec_VI} we give our conclusions, summarizing our results and their relevant aspects.

    Additionally, two appendices, Appendix \ref{A_1} and Appendix \ref{A_2}, containing explicit and detailed calculations concerning relevant parts of the article are included.

\section{Non-commutative GUP framework} \label{sec_II}

    The associative modified Heisenberg algebra we are going to analyze in this section is the $n$-dimensional generalization of the Kempf-Mangano-Mann algebra, extensively presented in \cite{Kempf:1994su}.
    In the one-dimensional scenario, the deformation of the canonical commutation relation is written as:
     \begin{equation} \label{II_1D CCR}
         [\hat{\mathbf{x}}, \hat{\mathbf{p}}]=i \hbar (1 + \beta \hat{\mathbf{p}}^2),
     \end{equation}
    where $\hat{\mathbf{x}}$ and $\hat{\mathbf{p}}$ are conjugate operators and $\beta$ is a positive deformation parameter, with proper dimensions.
    
    On momentum space wave functions $\psi(p)=\braket{p}{\psi}$, the operators can be represented as follows:
     \begin{align} 
         &\hat{\mathbf{p}}\ket{\psi} \rightarrow p \psi(p), \label{II_1D mom rep p}\\
         &\hat{\mathbf{x}}\ket{\psi} \rightarrow i \hbar (1+ \beta p^2) \partial_{p} \psi(p). \label{II_1D mom rep x}
     \end{align}
     It is straightforward to verify that this representation satisfies \eqref{II_1D CCR}.

     As exhaustively discussed in \cite{Kempf:1993bq, Kempf:1994su}, we can choose as domain of definition of these operators the Schwartz space $\mathcal{S}$, which is a dense domain on the Hilbert space of the theory $\mathcal{H}=\mathcal{L}^2(\mathbb{R}, dp/(1+\beta p^2)$.
     The modified Lebesgue measure assures the symmetry of the conjugate operators on the chosen domain $\mathcal{S}$.
     Nevertheless, a careful functional analysis shows that, while the $\hat{\mathbf{p}}$ operator is essentially self-adjoint on $\mathcal{S}$, the $\hat{\mathbf{x}}$ operator is not.
     Rather, the coordinate operator admits a one-parameter family of self-adjoint extensions.
     This feature is the mathematical sign of a departure from ordinary quantum mechanics.

     On a more physical ground, all of this is translated in the emergence of an absolute minimal uncertainty in the coordinate operator different from zero.
     
     This, in principle, should be true on the physical domain of the theory, which can be defined as the intersection of the domains of some fundamental operators \footnote{Here $\mathbf{\hat{x}}$ and $\hat{\mathbf{x}}^2$ refer to proper self-adjoint extensions of these operators.}:
      \begin{equation}
          \mathcal{D}_{phys}:=\mathcal{D}_{\mathbf{\hat{x}}} \cap \mathcal{D}_{\hat{\mathbf{x}}^2} \cap \mathcal{D}_{\mathbf{\hat{p}}} \cap \mathcal{D}_{\mathbf{\hat{p}^2}} \cap \mathcal{D}_{\comm{\mathbf{\hat{x}}_\lambda}{\mathbf{\hat{p}}}}.
      \end{equation}

    Nevertheless, this condition can be weakened, imposing that the minimization procedure holds for all the states belonging to the domain :
     \begin{equation}
          \mathcal{D}_{min}:=\mathcal{D}_{\mathbf{\hat{x}}_{\lambda}} \cap \mathcal{D}_{\hat{\mathbf{p}}}.
    \end{equation}
    
    Indeed, it can be shown that this condition is enough to define the uncertainties relations for a wave function (see \cite{Moretti:2013cma, Segreto:2022clx}.
    
    These states cannot be considered physical states in a strong way, as those which belong to $\mathcal{D}_{phys}$, but they can be considered physical states in a more "kinematic" sense.
    
    Throughout the paper, we will make reference to this last definition.
    
    The presence of this absolute minimum in the physical quantity $\Delta\hat{\mathbf{x}}$ is easily interpreted as a limit in the possibility of arbitrarily localizing objects in the configuration space of the theory.
    In light of this, the physical meaning of the coordinate representation, which is a point-wise concept, is lost.
    The eigenstates of the $\hat{\mathbf{x}}$ operator do exist but they have only a formal value as well as the basis they form.

    In this scenario, in order to recover information on localization in the configuration space, those states which realize the absolute minimal uncertainty in the coordinate operator can be used as a new basis.
    These states are called "maximally localized states" and, in the case in which the $\hat{\mathbf{x}}$ operator is the position operator, the new representation they give birth to is known as "quasi-position representation".

    The generalisation of this algebra to the $n$-dimensional case can be achieved by assuming the commutation relation \eqref{II_1D CCR} for every couple of conjugate variables:
     \begin{equation}
         [\hat{\mathbf{x}}_i, \hat{\mathbf{p}}_j]=i \hbar \delta_{ij}(1+\beta \hat{\vec{\mathbf{p}}}^2),
     \end{equation}
    where now the momentum on the right-hand side is the total momentum.

    We require that:
     \begin{equation}
         [\hat{\mathbf{p}}_i, \hat{\mathbf{p}}_j]=0.
     \end{equation}
   In this way, momentum representation is well-defined and we can generalize the representations \eqref{II_1D mom rep p}-\eqref{II_1D mom rep x} of the conjugate operators:
    \begin{align}
          &\hat{\mathbf{p}}_i\ket{\psi} \rightarrow p_i \psi(p), \\
          &\hat{\mathbf{x}}_i\ket{\psi} \rightarrow i \hbar (1+ \beta \vec{p}^{\; 2}) \partial_{p_i} \psi(p).  
    \end{align}
    
   By exploiting these representations it is possible to derive explicitly the commutation relations between coordinate operators:
    \begin{equation} \label{II_nD CR x op}
        [\hat{\mathbf{x}}_i, \hat{\mathbf{x}}_j]=2 i \hbar \beta (\hat{\mathbf{p}}_i \hat{\mathbf{x}}_j-\hat{\mathbf{p}}_j \hat{\mathbf{x}}_i).
    \end{equation}

   From \eqref{II_nD CR x op} it is clear that in this framework the $\hat{\mathbf{x}}$ operators do not commute between each other.
   This relevant feature of the theory can be interpreted as the emergence of a "non-commutative geometry" in the configuration space, fixed by the non-trivial non-commutativity of the coordinate operators.

   By introducing the operator:
    \begin{equation}
        \hat{\mathbf{L}}_{ij}=\frac{1}{1+\beta \hat{\vec{\mathbf{p}}}^2}( \hat{\mathbf{x}}_i \hat{\mathbf{p}}_j- \hat{\mathbf{x}}_{j} \hat{\mathbf{p}}_{i}),
    \end{equation}
   we can rewrite \eqref{II_nD CR x op} as follows:
     \begin{equation} \label{II_nD CR x op with L}
        [\hat{\mathbf{x}}_i, \hat{\mathbf{x}}_j]=-2 i \hbar \beta \left(1+\beta \hat{\vec{\mathbf{p}}}^2\right) \hat{\mathbf{L}}_{ij}.
    \end{equation}
 
    The operator $\hat{\mathbf{L}}_{ij}$ is the generator of \textit{n}-dimensional rotations in this quantum theory or, in other words, is the generalisation of ordinary orbital angular momentum operator, as it is clear from its momentum representation.

    Given that, it appears evident how the $n$-dimensional modified Heisenberg algebra we are dealing with preserves rotational symmetry, while the translation invariance is lost.
    This entails that the translation group is not defined in the usual sense in this quantum framework.

    The functional analysis of the operators in the $n$-dimensional case, even if more involved, is basically the same as the one-dimensional case.

    The Hilbert space of the theory is the space $\mathcal{H}=\mathcal{L}^2\left(\mathbb{R}^n, dp/(1+\beta \vec{p}^{\; 2})\right)$.
    The operators $\hat{\mathbf{x}}_i$ and $\hat{\mathbf{p}}_i$ can be defined on the Schwartz space $\mathcal{S}$, which is dense in $\mathcal{H}$.
    Here the  $\hat{\mathbf{p}}_i$ operators results to be symmetric and essentially self-adjoint.
    The formal eigenstates of the total momentum operator are $n$-dimensional Dirac deltas and for consistency we define their inner product as follows:
     \begin{equation} \label{II_Dirac_deltas_def}
       \braket{p}{p'}=(1+\beta \vec{p}^{\;2})\delta^{n}(p-p').
     \end{equation}

    On the other hand, the coordinate operators in every direction are symmetric operators on $\mathcal{S}$ but they are not essentially self-adjoint.
    Once again, each of them admits a one-parameter family of self-adjoint extensions.

    \noindent
    This fact suggests, by analogy with the one-dimensional case, that a limit in localization in configuration space arises in every direction.
    Nevertheless, in order to properly understand the structure of this quantum theory, it is important to recognize the relevant role played by the non-commutativity in the configuration space, which adds another challenge concerning objects localization.

    \section{Maximally localized states} \label{sec_III}

     We can use the \textit{squeezed states} method \cite{Kempf:1994su, Detournay:2002fq} to identify the maximally localized states of theory and the corresponding uncertainty in the configuration variables, for every direction.
     In the case of these uncertainties being different from zero, we can state that there is a limit in localization in the configuration space, i.e. minimal structures emerge within this framework.

     In this context by "squeezed states" we are referring to those states which saturate the uncertainty relation between conjugate operators:
      \begin{equation} \label{III_GUP relation}
          \Delta \hat{\mathbf{x}}_i \Delta \hat{\mathbf{p}}_j \! \geq \frac{\hbar}{2} \abs{\expval{[\hat{\mathbf{x}}_i, \hat{\mathbf{p}}_j]}} \! \rightarrow \! \Delta \hat{\mathbf{x}}_i \Delta \hat{\mathbf{p}}_j = \frac{\hbar}{2} \abs{\expval{[\hat{\mathbf{x}}_i, \hat{\mathbf{p}}_j]}}.
      \end{equation}
     
    Following \cite{Kempf:1994su, Detournay:2002fq}, from \eqref{III_GUP relation} it is possible to obtain a first order differential equation:
     \begin{align} \label{III_sq states diff eq}
         &\left[\hat{\mathbf{x}}_i-\xi_i+ i \hbar \delta_{ij} \Lambda (\hat{\mathbf{p}}_j-\eta_j)\right]\ket{\Psi_{\Lambda}^{i}}=0  \Rightarrow  \nonumber \\
         & \left[i \hbar (1+ \beta \vec{p}^{\; 2})\partial_{p_i} \! - \! \xi_i +i \hbar \delta_{ij} \Lambda (p_{j}-\eta_{j})\right]  \! \Psi_{\Lambda}^{i}(p)\!=\!0,
     \end{align}
    where we have set the following identities: 
     \begin{align}
         &\xi_i:=\expval{\hat{\mathbf{x}}_i}, \\ 
         &\eta_j:=\expval{\hat{\mathbf{p}}_j}, \\
         &\Lambda:=\frac{\expval{[\hat{\mathbf{x}}_i, \hat{\mathbf{p}}_j]}}{2 (\Delta \hat{\mathbf{p}}_j)^2}.
     \end{align}

   The solutions of \eqref{III_sq states diff eq} are exactly the squeezed states of theory if and only if they satisfy some specific conditions which are necessary to guarantee the connection between the differential equation \eqref{III_sq states diff eq} and relation \eqref{III_GUP relation} and the validity of the previous identities.
   An extensive discussion can be found in \cite{Detournay:2002fq} and in \cite{Segreto:2022clx}.

   In the $n$-dimensional case we are analyzing we can write $n$ differential equations \eqref{III_sq states diff eq}, one for every couple of conjugate variables.
   Let us call $\hat{\mathcal{A}}^{i}_{\Lambda}$ the operator which annihilates the state $\Psi_{\Lambda}^{i}$ in \eqref{III_sq states diff eq}.
   
   The first thing to notice is that:
    \begin{equation}
        [\hat{\mathcal{A}}^{i}_{\Lambda},\hat{\mathcal{A}}^{j}_{\Lambda}]= [\hat{\mathbf{x}}_i, \hat{\mathbf{x}}_j] \neq 0 \quad \text{if \;\;} i \neq j.
    \end{equation}
    From this simple commutation relation we can affirm that in this theory is not possible in general to find states that are maximally localized in different directions at the same time.
    This aspect of the theory is, as expected, the first consequence of the presence of a non-commutativity in configuration space: we can maximally localize an object in any direction, at the expense of de-localize it of a certain amount in all the other ones. 

    If we solve \eqref{III_sq states diff eq} for the generic $i$-th direction we obtain the following wave function \footnote{Here we solved the equation for $\eta_i=0$ since numerical computations show that the uncertainty in the coordinate operator is minimized by this choice. This can be further supported by the analysis of the one-dimensional case in \cite{Kempf:1994su}.}:
     \begin{align} \label{III_max loc state}
         \Psi^{i}_{\Lambda}(p)=&\mathcal{C} \; g\left(p_{\{j \neq i\}}\right) (1+ \beta p_{r}p_{s}\delta^{rs})^{-\frac{\Lambda}{2\beta}} \\
         &\times \exp{-i \xi_i \frac{\tan[-1](\frac{\sqrt{\beta}p_{i}}{\sqrt{1+\beta p_{r}p_{s}\delta^{rs}|_{r,s \neq i }}})}{\hbar \sqrt{\beta} \sqrt{1+\beta p_{r}p_{s}\delta^{rs}|_{r,s \neq i }}}},
     \end{align}
    where $g\left(p_{\{j \neq i\}}\right)$ is a not-specified function of momenta different from $p_i$, $\mathcal{C}$ is the normalization constant and the indices $r,s$ run over all the momenta, including $p_i$, except when differently specified.

    From expression \eqref{III_max loc state} it is clear that the final form of the maximally localized states and the respective uncertainty in $\hat{\mathbf{x}}_i$, will depend not only on the parameter $\Lambda$ but also on the function $g\left(p_{\{j \neq i\}}\right)$.

    Since we need to deal whit wave functions for which all the uncertainties relations hold and are well-defined, it is clear that this $g\left(p_{\{j \neq i\}}\right)$ function cannot be completely arbitrary.

    In order to fulfill our requests, the state \eqref{III_max loc state} have to be a normalizable state belonging to the domain of all the $\hat{\mathbf{p}}_k$ and $\hat{\mathbf{x}}_k$ operators (see above).
    Imposing these conditions entails some integrability requirements and accordingly some constraints over the possible values of $\Lambda$.
    More specifically, the bounds on $\Lambda$ are produced by the explicit integration over the $p_i$ variable, that we can perform in any case since the function $g\left(p_{\{j \neq i\}}\right)$ does not depend on $p_i$.

     The conditions the state has to meet are the following:
     \begin{itemize}
         \item normalizability:
                \begin{align}
                       &\int_{\mathbb{R}^{n-1}} \!\!\!\!\!\! d^{n-1}p\, (1+p_l p_m \delta^{lm})^{-\frac{1}{2}-\Lambda}\abs{g\left(p_{\{j \}}\right)}^2 < + \infty  \nonumber  \\
                      & \text{with} \quad \Lambda > -\frac{1}{2} 
                \end{align} 
          \item belonging to the domain of $\hat{\mathbf{p}}_i$: 
                \begin{align}
                       &\int_{\mathbb{R}^{n-1}} \!\!\!\!\!\! d^{n-1}p\, (1+ p_l p_m \delta^{lm})^{\frac{1}{2}-\Lambda}\abs{g\left(p_{\{j \}}\right)}^2 < + \infty  \nonumber  \\
                       &\text{with} \quad \Lambda > \frac{1}{2}
                \end{align} 
           \item belonging to the domain of $\hat{\mathbf{p}}_{j \neq i}$: 
                \begin{align}
                       &\int_{\mathbb{R}^{n-1}} \!\!\!\!\!\!\!\!\! d^{n-1}p\,  p_{j}^2 (1+ p_l p_m \delta^{lm})^{-\frac{1}{2}-\Lambda}\abs{g\left(p_{\{j \}}\right)}^2 \!\!< \! + \infty  \nonumber \\
                        & \text{with} \quad \Lambda > -\frac{1}{2}
                \end{align}   
            \item belonging to the domain of $\hat{\mathbf{x}}_{i}$: 
                \begin{align}
                       &\int_{\mathbb{R}^{n-1}} \!\!\!\!\!\! d^{n-1}p\, (1+ p_l p_m \delta^{lm})^{-\frac{1}{2}}\left[(1+p_l p_m \delta^{lm})\Lambda^2  \right.\\
                       &\left. + (2\Lambda -1)\left(\xi_{i}\right)^2 \right]\abs{g\left(p_{\{j \}}\right)}^2 < + \infty  \nonumber \\
                       &\text{with} \quad \Lambda > \frac{1}{2}
                \end{align} 
               \item belonging to the domain of $\hat{\mathbf{x}}_{j \neq i}$ \footnote{This condition was verified numerically.}: 
                 \begin{align}
                       &\int_{\mathbb{R}^{n}} \!\!\!\!\!\! d^{n}p\, \abs{i \hbar (1+p_{r}p_{s}\delta^{rs} \partial_{p_j} \Psi^{i}_{\Lambda}(p)}^2 < + \infty  \nonumber \\
                       &\text{with} \quad \Lambda > \frac{3}{2}
                \end{align} 
        \end{itemize}

    As it should be clear, in the previous expressions the indices $j,l,m \neq i$ and we have recast everything in the "natural" units of the theory, that is $[\Lambda]=\beta$ and $[p]=1/\sqrt{\beta}$.
    
    We pause a moment to clarify the choice of domains with respect to which we determined the previous conditions.
    For the operators $\hat{\mathbf{p}}_i$ and $\hat{\mathbf{p}}_{j \neq i}$ we considered the only extension of $\mathcal{S}$ in which they result to be self-adjoint (see e.g. \cite{Segreto:2022clx}).
    Concerning the position operators - which we remind are not essentially self-adjoint - we considered the domain of the adjoint operators.
    This is indeed a suitable choice since all the self-adjoint extensions of the starting domain $\mathcal{S}$ are contained in the domain of the $\hat{\mathbf{x}}^{\dagger}$ operators.
    Therefore, even though we are not particularly concerned with ensuring that the maximally localized states belong to the domain of adjoint operators, due to the aforementioned inclusion relation, imposing belonging in these domains establishes a minimal condition that these states must still satisfy. The specific belonging to "smaller" self-adjoint domains may at most impose more stringent conditions.

    Now, in order to understand the role played by this function in determining the minimum of the uncertainty of the $\hat{\mathbf{x}}_i$ operator, we can explicitly derive the expression of $\Delta\hat{\mathbf{x}}^{i}_{\Lambda}$ for a generic function $g\left(p_{\{j \neq i\}}\right)$, which results to be:
    \begin{align} \label{III_general_g_uncert}
     \Delta\hat{\mathbf{x}}^{i}_{\Lambda}\!\!=\!\!& \left[\!\frac{ \bigintsss_{\mathbb{R}^{n-1}} \! d^{n-1}p \, \abs{g\left(p_{\{j \}}\right)}^2 (1\!+\!p_l p_m\delta^{lm})^{\frac{1}{2}-\Lambda}}{ \bigintsss_{\mathbb{R}^{n-1}} \! d^{n-1}p \, \abs{g\left(p_{\{j\}}\right)}^2 (1\!+\!p_l p_m\delta^{lm})^{-\frac{1}{2}-\Lambda}} \nonumber \right.\\
     &\left. \times \left(\frac{\Lambda^2}{2 \Lambda -1}\right) \right]^{\frac{1}{2}}  \hbar \sqrt{\beta}
    \end{align}
    where again the indices $j,l,m \neq i$ and  we have carried out the integration in the $p_i$ variable.

    From basic properties of the Lebesgue integral, we can affirm that the ratio of the two integrals in \eqref{III_general_g_uncert} is always greater than one.
    Indeed the first integrand is always greater than the second one on the whole $\mathbb{R}^{n-1}$ space, except at the origin, where the two are equals.
    
    On this ground we can write:
     \begin{equation}
          \Delta\hat{\mathbf{x}}^{i}_{\Lambda} > \frac{\Lambda^2}{2 \Lambda -1} \hbar \sqrt{\beta} 
     \end{equation}
    Since the right-hand side of the previous expression reaches its minimum for $\Lambda= 1$ (in $\beta$ units), finally we deduce that  $\Delta\hat{\mathbf{x}}^{i}_{\Lambda}>1$ (in $\hbar \sqrt{\beta}$ units).
    We notice that the obtained lower bound is exactly the minimal uncertainty found in \cite{Kempf:1994su} for the one-dimensional case.

    The existence of the bound itself for $ \Delta\hat{\mathbf{x}}^{i}_{\Lambda}$ suggests that, in principle, it should be possible to model the function $g(p_{\{j\}})$ in order to make the uncertainty arbitrarily close to one.

    Thinking of the $g-$functions as weights for integral, it is not hard to understand that this is exactly what happens when we select functions centered in the origin, with a gradually narrower profile.

    Indeed, by choosing this kind of functions, the main contribution for the two integrals comes from the region around the origin, where the two integrands are essentially the same.
    
    Therefore, in a similar scenario, while the ratio of the two integrals tends towards one, accordingly we will have:
     \begin{equation}
          \Lambda_{min} \to 1  \quad  \text{and} \quad \Delta\hat{\mathbf{x}}_i \to 1.
     \end{equation}

    Nevertheless, all those states which tend to realize this condition are not acceptable, since it is not possible to consistently define for them the uncertainties in all the other directions.
    Indeed, all the states for which $\Delta\hat{\mathbf{x}}_i \to 1$, realize this condition for $\Lambda_{min} \to 1$ and hence they violate the previous fundamental conditions that we have derived on $\Lambda$ in order to deal with a proper wave function.
    In particular, it is the condition of belonging to the domain of the $\hat{\mathbf{x}}_{j \neq i}$ operators which is not fulfilled in the case of $\Lambda \to 1$.
    
    From a physical point of view, by analogy with ordinary quantum mechanics, we could say that these states, while almost maximally localized in one direction, tend to be completely delocalized in all the other ones, as it could be reasonably expected due to the presence of non-commutativity. 
    
    From all these considerations it should be now clear that - if an absolute minimum in the uncertainty of the $i$-th coordinate operator exists - the $g$-function realizing this condition must be identified without contravening the constraints we have established.

    This could be done by means of a constrained variational principle (see \cite{Detournay:2002fq, Segreto:2022clx}).
    Nevertheless, we will not carry out this task further, here.
    We simply observe, for example, that a constant $g$-function, while not guaranteed to be the one leading to the minimum configuration, satisfies all the requirements and, as it will be clearer from what follows, it plays somehow a special role.

    The point is that, through this procedure, we could be able to obtain states maximally localized only along one direction, containing only information about that specific direction.
    This information is carried by the $\xi_i$ parameter, which is the expectation value of the $i$-th coordinate operator for the considered wave function.

    These states alone cannot help us to map the whole configuration space and hence to recover in a satisfactory way information on position.

    To pursue this task we start from the following consideration: even if it is true that in general the $\hat{\mathcal{A}}^{i}_{\Lambda}$ operators do not commute each other, there is a particular case that represents an exception.
    Indeed, if the $\xi_i$ parameters are zero the $\hat{\mathcal{A}}^{i}_{\Lambda}$ operators do commute and this means that they have a common eigenstates basis.
    
    From the wave function in \eqref{III_max loc state} it appears evident that this family of states is represented by:
     \begin{equation}
         \Phi_{\Lambda}(p)=\mathcal{N}(1+\beta p_{s}p_{r}\delta^{sr})^{-\frac{\Lambda}{2\beta}}.
     \end{equation}

     This object is a rotational invariant state in momentum space and its expectation value of position in configuration space is zero.

     We can easily compute the normalization constant $\mathcal{N}$ using hyperspheric coordinates: 
      \begin{equation} \label{III_max_loc_rot_inv}
           \Phi_{\Lambda}(p)\!=\!\!\sqrt{\left(\frac{\beta}{\pi}\right)^{n/2}\!\!\!\!\!\frac{\Gamma\left[1+\frac{\Lambda}{\beta}\right]}{\Gamma\left[1-\frac{n}{2}+\frac{\Lambda}{\beta}\right]}}(1+\beta p_{r}p_{s}\delta^{rs})^{-\frac{\Lambda}{2\beta}}\!.  
      \end{equation}
    
    With the same technique, we can also obtain the uncertainties in the coordinate operators, which are the same in every direction:
     \begin{align} \label{III_unc_pos_max_state_rot_inv}
         &(\Delta \hat{\mathbf{x}}_i)^2_{\Lambda}\!=\!\! - \hbar^2 \!\! \!\int_{\mathbb{R}^n} \!\!\!\!\! d^{n}p \; \Phi^{*}_{\Lambda}(p) \partial_{p_i} \! \left[ \left(1+\beta p_r p_s\delta^{rs} \right) \partial_{p_i}\Phi_{\Lambda}(p)\right] \!  \nonumber \\
         &=\hbar^2 \left(\frac{\beta}{\pi}\right)^{n/2}  \frac{\Gamma\left[1+\frac{\Lambda}{\beta}\right]}{\Gamma\left[1-\frac{n}{2}+\frac{\Lambda}{\beta}\right]}  \int  d\Omega_{n-1} \int_{\mathbb{R}^+} dr \; r^{n-2}  \nonumber \\
         &\!\! \times \!\! \int_{\mathbb{R}}\!\! dp_{i}\! \left(1+\beta r^2 \!\! +p^2_{i} \right)^{-\frac{\Lambda}{\beta}-1} \! (1+\beta r^2\! -\!p^{2}_{i}(-\beta+\Lambda) )\Lambda  \nonumber \\
         &=\frac{\Lambda^2}{(2 \Lambda-n )} \; \hbar^2 \beta , \quad \quad  \Lambda> \frac{n}{2} , \;  n \in \mathbb{N},
     \end{align}
    where $r=\sum_{j\neq i} p^2_{j}$, $\Omega_{n-1}$ is the $(n-1)$-dimensional solid angle and, in the result, we are using again the natural units suggested by theory, i.e. $[x_{i}]=\hbar \sqrt{\beta}$ and $[\Lambda]=\beta$.
    
    In order to reach the absolute minimum value of the quantity $(\Delta \hat{\mathbf{x}}_i)^2_{\Lambda}$ for the wave function \eqref{III_max_loc_rot_inv}, we need to minimize \eqref{III_unc_pos_max_state_rot_inv} with respect to $\Lambda$.
    This operation leads to the following result:
     \begin{equation} \label{III_unc_pos_min_rot_inv}
         \Lambda_{min}= n  \Rightarrow \Delta \hat{\mathbf{x}}_i^{min}=\sqrt{n} \; \hbar \sqrt{\beta}.
     \end{equation}

     In the end we can write the desired expression for the wave function \eqref{III_max_loc_rot_inv}:
      \begin{equation} \label{III_max_loc_state_rot_inv_final}
          \Phi^{ml}(p)=\sqrt{\left(\frac{\beta}{\pi}\right)^{n/2}\frac{\Gamma\left[1+n\right]}{\Gamma\left[1+\frac{n}{2}\right]}}(1+\beta \; p_{r}p_{s}\delta^{rs})^{-\frac{n}{2}}. 
      \end{equation}

     We stress the fact that, even if the second line integral in \eqref{III_unc_pos_max_state_rot_inv} is defined only for $n>1$, the final result is valid $\forall n \in \mathbb{N}$.
     This is confirmed by expressions \eqref{III_unc_pos_min_rot_inv} and \eqref{III_max_loc_state_rot_inv_final}, which, for $n=1$, are exactly the maximally localized state (with $\xi=0$) and the relative uncertainty found in \cite{Kempf:1994su} for the one-dimensional case.

     The state \eqref{III_max_loc_state_rot_inv_final} appears then to be a special state for the theory, since it is the only state - at least according to the squeezed state method - which can be maximally localized by the same amount, simultaneously in every direction.

     \section{Quasi-position basis} \label{sec_IV}

     As stated in the previous section, maximally localized states along one direction are not enough to recover information on position in configuration space.
     This holds also for the special state \eqref{III_max_loc_state_rot_inv_final}, which cannot be used as a basis to construct a new representation, since there is no parameter that can promoted to a variable.
     Nevertheless, the obtained wave function preserves its privileged role as a function able to minimize by the same amount at the same time the uncertainties in the coordinate operators.
     Hence, to reintroduce the parameter $\xi_i$ and construct a new basis to map the configuration space, following the path suggested in \cite{Kempf:1994su} for the quasi-position representation (for a discussion of this representation see \cite{Bosso:2020aqm, Bosso:2021xsn}, we propose to "translate" in every direction the state \eqref{III_max_loc_state_rot_inv_final}, so that we can obtain a new state for which holds the following:
      \begin{equation} \label{IV_def_transl}
          \frac{\bra{\Phi^{ml}_{T}}\hat{\mathbf{x}}_k \ket{\Phi^{ml}_{T}}}{\braket{\Phi^{ml}_{T}}}=\xi_k, \quad \forall k.
      \end{equation}

    The latter expression has to be taken as a definition for what we mean in this scenario for a translation, since, strictly speaking, the translation group in the usual sense is not defined in this kind of theories.

    The new wave function which allows us to achieve the validity of the expression \eqref{IV_def_transl} is suggested from the form of the state \eqref{III_max loc state}:
     \begin{align} \label{IV_quasi_pos_basis}
         &\Phi^{ml}_{T}(p)=\sqrt{\left(\frac{\beta}{\pi}\right)^{n/2}\frac{\Gamma\left[1+n\right]}{\Gamma\left[1+\frac{n}{2}\right]}}(1+\beta \; p_{r}p_{s}\delta^{rs})^{-\frac{n}{2}} \nonumber \\
         &\times \prod_{k=1}^{n} \exp{-i \xi_k \frac{\arctan(\frac{\sqrt{\beta}p_{k}}{\sqrt{1+\beta p_{r}p_{s}\delta?{rs}|_{r,s\neq k}}})}{\hbar \sqrt{\beta} \sqrt{1+\beta p_{r}p_{s}\delta^{rs}|_{r,s\neq k}}}}.
     \end{align}
    The normalization constant is unchanged since we have only added complex exponential factors to the state \eqref{III_max_loc_state_rot_inv_final} and it is not difficult to show that the conditions \eqref{IV_def_transl} are satisfied.
    The uncertainties in the coordinate operators can be found after performing several $n$-dimensional integrals by means of hyperspheric coordinates and the obtained expression is the same for any direction.
    The details of this calculation can be found in the appendix.
    In the end we can write:
    \begin{align} \label{IV_unc_pos_nc_basis}
        (&\Delta{\hat{\mathbf{x}}_i})^2=\frac{1}{2}\left\{ 2n+\frac{2+4n}{(n-1)(4n^2-1)}\sum_{k\neq i} \xi^2_{k}  \right. \nonumber \\
        &\left. \!\! + \frac{n!}{\Gamma[3/2+n]}\sqrt{\pi}\biggl(\alpha(n)+\theta(n)\pi^2\biggr) \sum_{k\neq i} \xi^2_{k} \right\} \hbar^2 \beta,
    \end{align}
   where $\alpha(n)$ and $\theta(n)$ are respectively a real negative function and a real positive function depending on the dimension $n$, for which there is no closed form.
   A table of their values can be found in the appendix \ref{A_1}.

   The most relevant fact to notice is that the uncertainties in the coordinate operators are now dependent on the $\xi_i$ parameters, i.e. the expectation values on the configuration variables $x_i$. 
   Therefore, by using states \eqref{IV_quasi_pos_basis} as a basis, the possibility of localizing an object in the configuration space and the limit to this procedure will depend on \textit{where} we want to localize.
   This means that with respect to the localizing properties, the configuration space of the theory appears to be somehow "inhomogeneous".
   In particular, from \eqref{IV_unc_pos_nc_basis}, it is clear that, once we have fixed an origin in our configuration space, away from this point the localization process becomes increasingly worse. 
   Accordingly, we can strictly talk about localized objects only around the fixed origin. 

   Now that we have constructed our basis we can define a map from momentum space to quasi-position space:
    \begin{align} \label{IV_map_mom_to_quasipos}
        \tilde{\psi}(\xi):=&\int_{\mathbb{R}^n} \frac{d^n p}{(1+\beta p_{r}p_{s}\delta^{rs})}\braket{\Phi_{T}^{ml}}{p}\braket{p}{\psi} \nonumber \\
        =&\int_{\mathbb{R}^n} \frac{d^n p}{(1+\beta p_{r}p_{s}\delta^{rs})}\overline{\Phi_{T}^{ml}(p)} \psi(p).
    \end{align}
  This map can be properly inverted, obtaining a transformation from quasi-position space to momentum space:
   \begin{align}
       &\psi{(p)}:=\frac{(1+\beta p_{r}p_{s}\delta^{rs})^{\frac{n}{2}+1}}{\sqrt{2 \pi}^{n}}\abs{\mathcal{J}_{n}\left[V(p)\right]}\int_{\mathbb{R}^n} \!\!\! d^n \xi \; \psi(\xi) \\
       & \times \prod_{k=1}^{n} \exp{-i \xi_k \frac{\arctan(\frac{\sqrt{\beta}p_{k}}{\sqrt{1+\beta p_{r}p_{s}\delta^{rs}|_{r,s\neq k}}})}{\hbar \sqrt{\beta} \sqrt{1+\beta p_{r}p_{s}\delta^{rs}|_{r,s\neq k}}}},
   \end{align}
  where $V(p)$ is a vector field properly defined and $\mathcal{J}$ is the associated Jacobian matrix.
   
   How we have constructed the inverse transform guarantees the unitarity of the transform itself.
   
   The details of the derivation of the inverse transform and all the necessary definitions can be found in the appendix \ref{A_2}.

\section{Non-commutative GUP quantization of Bianchi I model} \label{sec_V}

  In this section, we are going to discuss the quantization procedure of a cosmological model, namely the Bianchi I model, within the framework we have developed so far.
  As it is widely known, one-particle quantization schemes are well-suited for the cosmological context.
  This is true since space-times considered in cosmology are usually homogeneous space-times, therefore the gravitational field we are dealing with has a finite number of degrees of freedom.
  In this scenario, a GUP quantization procedure is potentially able to provide the theory with some new relevant features, such as the emergence of minimal structures and different localization properties due to the presence of non-commutativity in configuration space, which can modify in a meaningful way the dynamics of the Universe and its behavior.

  In order to apply rigorously the machinery we have developed to our cosmological model, we need to work in a quantum formulation where the probabilistic interpretation of the wave function of the system is well-posed and meaningful and where the dynamics is dictated by a true Schr\"odinger equation.

  This is a delicate issue in quantum cosmology, which can be addressed in different ways.
  In this paper we will adopt the Vilenkin approach \cite{Vilenkin:1988yd}.
  In Vilenkin's idea, in order to achieve the above configuration, we need to separate the system, by means of its degrees of freedom, in a semi-classical part and a full quantum part.
  As we will see, it is the presence of a part of the system treated semi-classically that allows us to obtain the desired dynamics for the quantum part.

  The conditions that must be satisfied in order to have a consistent and well-defined procedure are extensively discussed in \cite{Vilenkin:1988yd}.
  
  Here we stress the fact that we need to assume that the quantum subsystem has a negligible effect on the semi-classical part.
  In this sense we can talk about a "small" quantum part of the system.
    
  Now, we are going to re-derive explicitly every step for the particular chosen case, working in momentum representation.
  As it should be clear from the previous sections, this is the natural choice within a GUP theory.

  The configurational variables and the respective conjugate momenta we are adopting are the so-called Misner variables $(\alpha, \gamma_{+},\gamma_{-})$ and $(p_{\alpha},p_{+},p_{-})$ \cite{Misner:1967uu}.
  These variables are able to diagonalize the kinetic part of the Hamiltonian of our system - and this is true for several Bianchi models - and it can be shown that they have a clear physical meaning.
  Indeed, the variable $\alpha$ is related to the isotropic volume of the Universe, while the variables $\gamma_{+}$ and $\gamma_{-}$ are known as \textit{anisotropies}, since they "measure" the deviation from the completely isotropic Universe.
  
  The Hamiltonian constraint describing a Bianchi I universe is the following:
   \begin{equation}
       \mathcal{H}= N(t) e^{-3 \alpha}\left(-p_{\alpha}^2+p_{+}^2+p_{-}^2\right)\approx 0,
   \end{equation}
  where $N(t)$ is the lapse function,  coming out from the $3+1$ reduction procedure of the metric \cite{Arnowitt:1962hi}, where we have set the speed of light in vacuum $c$ and the Einstein constant $\kappa$ is equal to one.

  For a consistency matter, as it will be clear later, we need to consider additionally a positive cosmological constant term in our Universe, hence rewriting the Hamiltonian constraint as:
   \begin{equation} \label{V_Ham_constraint}
       \mathcal{H}=N(t) e^{-3 \alpha}\left(-p_{\alpha}^2+p_{+}^2+p_{-}^2+ \Lambda e^{6 \alpha} \right)\approx 0.
   \end{equation}
  By inspecting \eqref{V_Ham_constraint} it is evident how the cosmological constant part acts as a potential for the system. 

  The constraint \eqref{V_Ham_constraint} has to be promoted to be a quantum operator:
   \begin{equation} \label{V_quantum_constraint}
       \hat{\mathbf{\mathcal{H}}}=\left(- \hat{\mathbf{p}}_{\alpha}^2+ \Lambda e^{6 \hat{\mathbf{\alpha}}}+\hat{\mathbf{p}}_{+}^2+\hat{\mathbf{p}}_{-}^2\right) \ket{\Psi}= 0,  
   \end{equation}
  where we can identify two parts:
   \begin{itemize}
       \item the semi-classical part $\hat{\mathcal{H}}_{0}=- \hat{\mathbf{p}}_{\alpha}^2+ \Lambda  e^{6 \hat{\mathbf{\alpha}}}$, which concerns essentially the volume of the Universe;
       \item the full quantum part $ \hat{\mathcal{H}}_{q}=\hat{\mathbf{p}}_{+}^2+\hat{\mathbf{p}}_{-}^2$, which is completely described by the anisotropies.
   \end{itemize}

  By considering a parameter $\lambda \propto \hbar$, we can express the "smallness condition" of the quantum subsystem with respect to the semi-classic part in a more precise way:
   \begin{equation}
       \hat{\mathcal{H}}_{q}/ \hat{\mathcal{H}}_{0}=o(\lambda).
   \end{equation}

  By following the reasoning line of \cite{Vilenkin:1988yd}, we can write the ansatz for the solution of \eqref{V_quantum_constraint} in momentum space:
   \begin{align}
       \Psi(p_{\alpha},p_{\pm}) &=  A(p_{\alpha})e^{i S(p_{\alpha})/ \hbar} \chi(p_{\alpha},p_{\pm}) \\
       &:= \Psi_0(p_{\alpha})\chi(p_{\alpha},p_{\pm}).
   \end{align}
  We can distinguish the $\Psi_0$ term, which is a WKB expansion of the semi-classical part with respect to the $\hbar$ parameter and the $\chi$ term containing information about the quantum behavior of the system.

  We will now set two different equations for the two different terms of the wave function of the Universe and we will solve them in order to obtain a complete description of the system, according to the method we are using.

  The wave function $\Psi_0$ will satisfy the equation:
   \begin{equation} \label{V_semicl_eq}
       \bra{p}\hat{\mathcal{H}}_0 \ket{\Psi_0}\!=\!\left(\!-p^2_{\alpha}+\Lambda e^{i 6 \hbar \partial_{p_{\alpha}}}\!\right) A(p_{\alpha})e^{i S(p_{\alpha})/\hbar}\!=\!0.
   \end{equation}

  In the spirit of the WKB expansion, being $\hbar$ the expansion parameter, we can solve \eqref{V_semicl_eq} at different orders \cite{Landau:1991wop}.
  Specifically, we will solve \eqref{V_semicl_eq} at the zeroth and first order in $\hbar$, obtaining in this way respectively a Hamilton-Jacobi equation, which will allow us to determine $S(p_{\alpha})$ and a continuity equation from which will be possible to derive $A(p_{\alpha}).$

  We will refer to the formal series expansion of the pseudo-differential operator in \eqref{V_semicl_eq} to identify the different orders in $\hbar$:
   \begin{equation} \label{V_pseudodiff_op}
       e^{i 6 \hbar \partial_{p_{\alpha}}}=\sum_{n=0}^{\infty}\frac{\left(i 6 \hbar \partial_{p_{\alpha}}\right)^n}{n!}.
   \end{equation}

  By letting \eqref{V_pseudodiff_op} acting on $\Psi_{0}$, at the zeroth order in $\hbar$, the equation \eqref{V_semicl_eq} will read as:
   \begin{align} \label{V_H-J_eq}
      &\left(\!\!- p_{\alpha}^2\! +\!\Lambda \sum_{n}^{\infty}\frac{1}{n!}(6 \; \partial_{p_{\alpha}} S(p_{\alpha}))^{n}\!\right) \! A(p_{\alpha})e^{i S(p_{\alpha})/\hbar}\!=\!0 \nonumber \\
      &- p_{\alpha}^2+ \Lambda e^{6 \partial_{p_{\alpha}} S(p_{\alpha})}\!=\!0,
   \end{align}
  from which it is straightforward to obtain:
   \begin{equation}
      S(p_{\alpha})=\frac{1}{6}p_{\alpha}\left( \ln(\frac{ p_{\alpha}^2}{\Lambda})-2\right)+c_{1}.
   \end{equation}

  The function that we have just determined, by working at the zeroth order in $\hbar$, is the classical action of the system only concerning the variables $(\alpha, p_{\alpha})$.
  
  On the other hand, at the first order in $\hbar$, considering the validity of \eqref{V_H-J_eq}, we can write the following differential equation for $A(p_{\alpha})$:
   \begin{equation} \label{V_cont_eq}
          (6 S')^2A'e^{6 S'}+6 A S''(e^{6 S'}-6S'-1)=0,
   \end{equation}
   where the symbol $'$ refers to the differentiation with respect to $p_{\alpha}$.
  The solution of \eqref{V_cont_eq} is:
   \begin{equation}
       A(p_{\alpha})=c_{2} \; e^{-\frac{\Lambda -  p^2_{\alpha}}{ p^2_{\alpha} \ln(\frac{ p^2_{\alpha}}{\Lambda})}}.
   \end{equation}
  Now that we have completely described the semi-classical part of the wave function of the Universe, we can turn our attention to the quantum part.
  The quantum piece of the wave function $\chi$ depends not only on the quantum variables $p_{\pm}$ but also on the semi-classical one $p_{\alpha}$. This means that, in order to obtain a correct description of the dynamics, we need to consider both the action of $\hat{\mathcal{H}}_q$ and $\hat{\mathcal{H}}_0$ on $\chi$.
  Working at the first order in $\hbar$ - which is the first order containing quantum information - and keeping in mind the validity of the equations \eqref{V_H-J_eq} and \eqref{V_cont_eq} for the semi-classical part, we arrive at the following equation for the quantum subsystem:
   \begin{equation} \label{V_quantum_dyanmics}
      6 i \hbar \Lambda e^{6 S'(p_{\alpha})}  \frac{\partial \chi (p_{\alpha}, p_{\pm})}{\partial p_{\alpha}}=-\mathcal{H}_q \chi(p_{\alpha}, p_{\pm}).
   \end{equation}

  Too obtain, in the end, a real Schr\"odinger equation from \eqref{V_quantum_dyanmics}, we need to exploit a series of relations which holds for the classical action $S(p_{\alpha})$ and the classical quantities related to this object.
  First of all, from the Hamilton-Jacobi construction, we know that $\partial_{p_{\alpha}} S= \alpha$ and we can rewrite \eqref{V_quantum_dyanmics} as:
   \begin{equation} \label{V_almost_SE}
      6 i \hbar\Lambda  e^{6 \alpha }\frac{\partial \chi (p_{\alpha}, p_{\pm})}{\partial p_{\alpha}}=-\mathcal{H}_q \chi(p_{\alpha}, p_{\pm}). 
   \end{equation}
  Secondly, we need to consider the classical Hamiltonian described by the variables $(\alpha, p_{\alpha})$:
   \begin{equation}
       \mathcal{H}_{cl}=c^3 N(t) e^{-3 \alpha}\left(\ p_{\alpha}^2+ \Lambda e^{6 \alpha} \right)\approx 0.
   \end{equation}
   By taking into account the respective Hamilton equations we can arrive at the following expression:
    \begin{equation} \label{V_semicl_rel_towards_SE}
        6 \Lambda e^{6 \alpha}=- \frac{\dot{p_{\alpha}}}{ N(t)}e^{3 \alpha}
    \end{equation}
  where the dot refers to the differentiation with respect to the coordinate time.

  Finally, putting together \eqref{V_almost_SE}-\eqref{V_semicl_rel_towards_SE}, we are able to write \eqref{V_quantum_dyanmics} as:
   \begin{align}  \label{V_final_SE}
       &i \hbar \dot{p_{\alpha}}\frac{\partial \chi (p_{\alpha}, p_{\pm})}{\partial p_{\alpha}}=N(t) e^{-3 \alpha} \mathcal{H}_q \chi(p_{\alpha}, p_{\pm}) \Rightarrow  \nonumber \\
       &i \hbar \frac{\partial \chi (p_{\alpha}, p_{\pm})}{\partial t}= N(t) e^{-3 \alpha}  \mathcal{H}_q \chi(p_{\alpha}, p_{\pm}),
   \end{align}
 which is the Schr\"odinger equation for the quantum dynamics we were looking for.

 We notice that in \eqref{V_final_SE} the lapse function $N(t)$ is present: this is a reflection of the fact that our theory is invariant under time-diffeomorphism, as it should be.
 Indeed, we can choose different times to follow the evolution of the system and consequently we will deal with different Hamiltonians.

 In our case,  we want the Hamiltonian of the quantum subsystem of the Universe to be the one of a free particle.
 This implies to set $N(t)$ to a precise form, specifically:
  \begin{equation}
      N(t)=e^{3 \alpha(t)}.
  \end{equation}
 Consequently, we are able to define our physical time $\tau$, which is essentially a WKB time, from the relation:
  \begin{equation}
     d \tau= e^{3 \alpha(t)}dt.
  \end{equation}
 In the end:
  \begin{equation} \label{V_final_final_SE}
      i \hbar \frac{\partial \chi (p_{\alpha}, p_{\pm})}{\partial \tau}=  \left(p^2_{+}+p^2_{-}\right) \chi(p_{\alpha}, p_{\pm})
  \end{equation}

 The general solution of \eqref{V_final_final_SE} is given by a superposition of free particle time-dependent wave functions.
 In the non-commutative GUP quantization scheme - which is the one we are adopting for the description of the quantum degrees of freedom of our Bianchi I Universe - these wave functions can be obtained by the map \eqref{IV_map_mom_to_quasipos}, transforming into quasi-position representation - or in this case, quasi-anisotropies - two dimensional Dirac deltas, defined as in \eqref{II_Dirac_deltas_def}.
 This allows us to write the general solution of \eqref{V_final_final_SE} as:
  \begin{align} \label{V_general_sol_SE}
      &\tilde{\chi}(\tilde{\gamma}_{+},\tilde{\gamma}_{-})=\int_{\mathbb{R}^2} \frac{dp_{+} dp_{-}}{\left(1+\beta \; (p_{+}^2+p_{-}^2)\right)^2}  \\ 
      &\times \exp{i \tilde{\gamma}_{+} \frac{\arctan(\frac{\sqrt{\beta}p_{+}}{\sqrt{1+\beta p_{-}^2}})}{\hbar \sqrt{\beta} \sqrt{1+\beta p_{-}^2}}} \nonumber \\
      & \times  \exp{i \tilde{\gamma}_{-} \frac{\arctan(\frac{\sqrt{\beta}p_{-}}{\sqrt{1+\beta p_{+}^2}})}{\hbar \sqrt{\beta} \sqrt{1+\beta p_{+}^2}}} \nonumber \\
      & \times \exp{-i  (p_{+}^2+p_{-}^2) \frac{(\tau-\tau_0)}{\hbar}}\phi_{0}(p_{+},p_{-}).
  \end{align}

 To explore explicitly the behavior of the system we need to choose a precise profile $\phi_{0}(p_{+},p_{-})$, which represents also the wave function of the system in momentum space at some initial time $\tau_0$.

 \begin{figure*}[!ht]
     \includegraphics[width=%
  	0.32\textwidth]{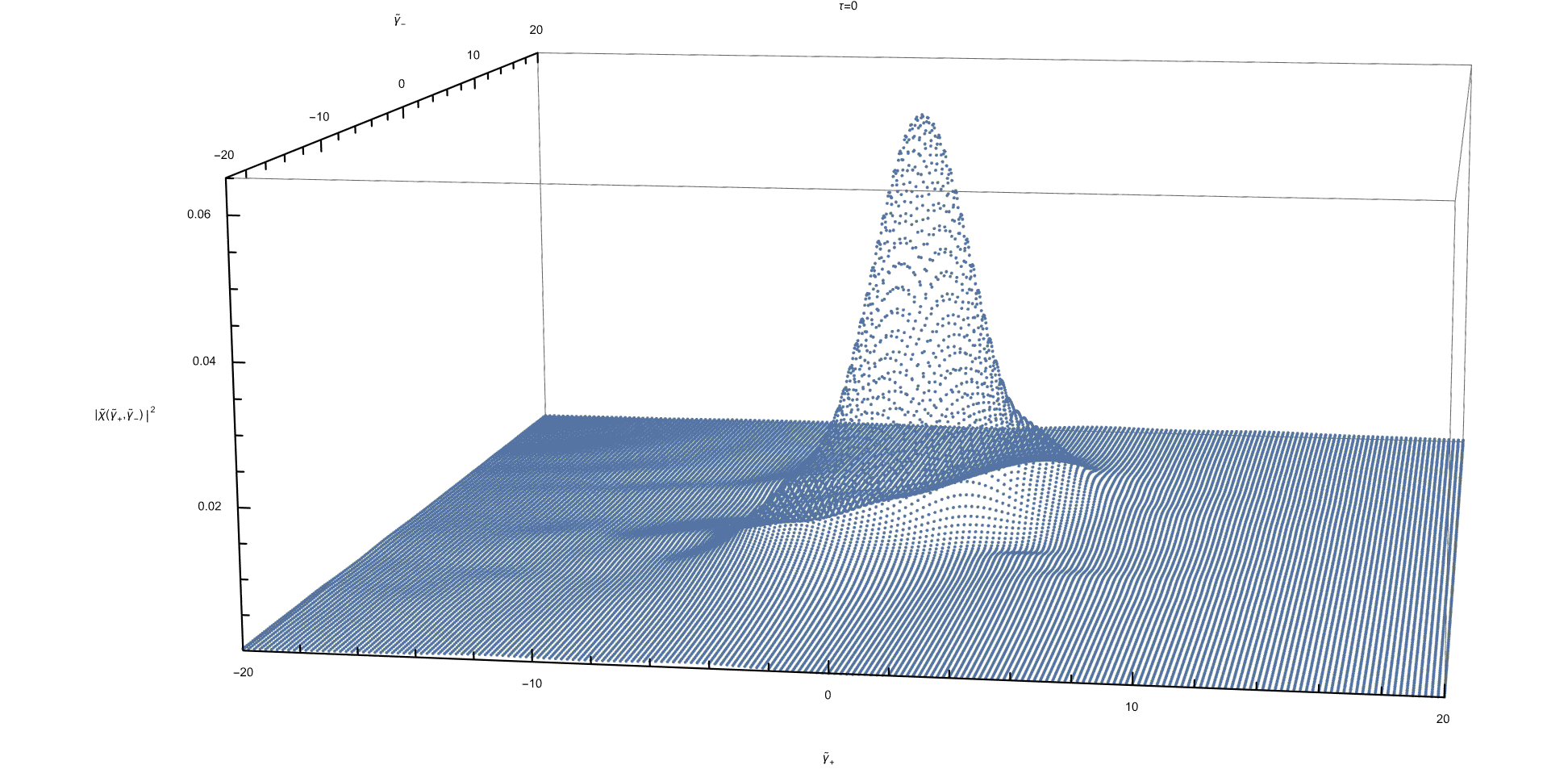}
     \includegraphics[width=%
  	0.32\textwidth]{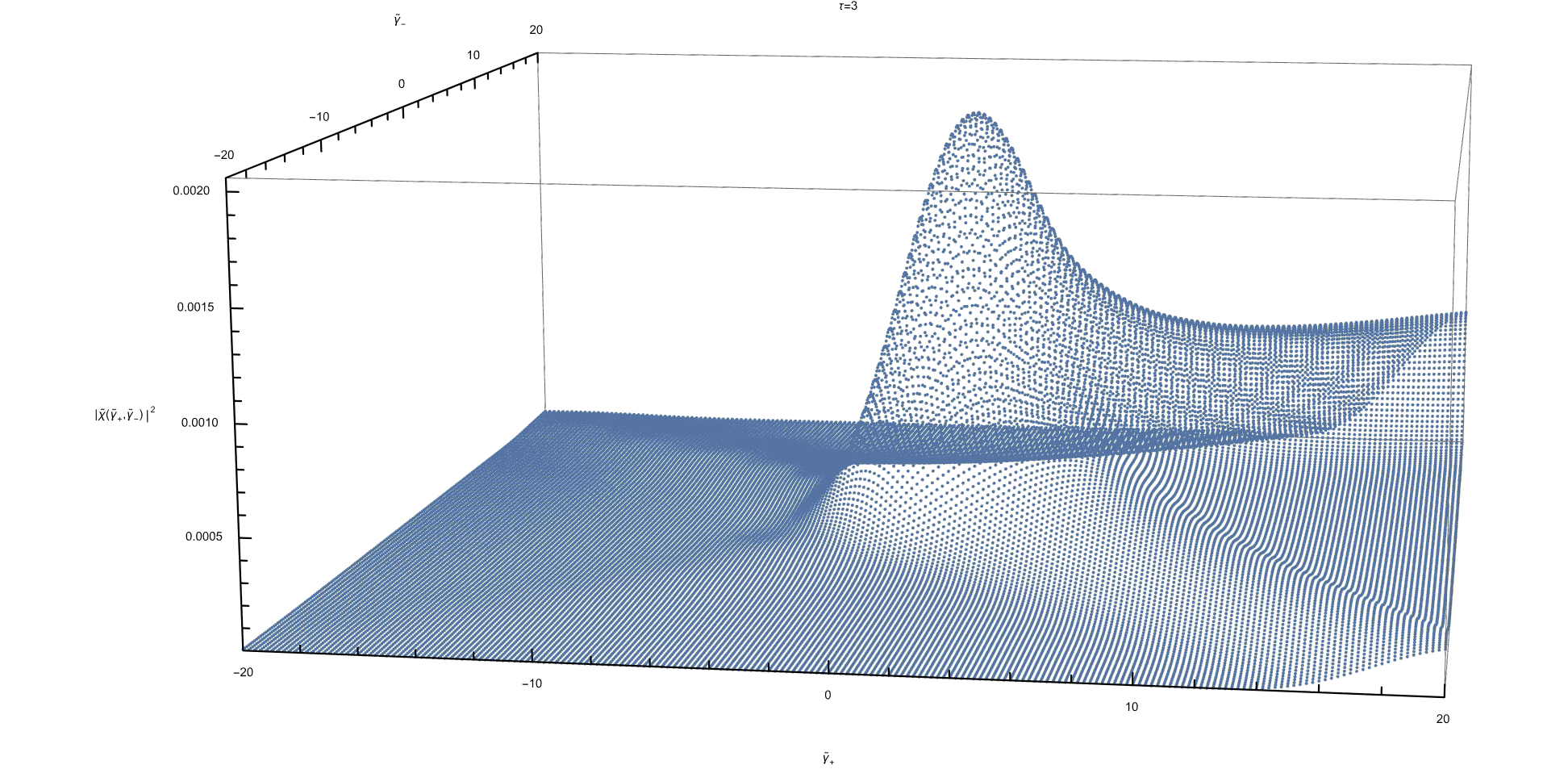}
     \includegraphics[width=% 
     0.32\textwidth]{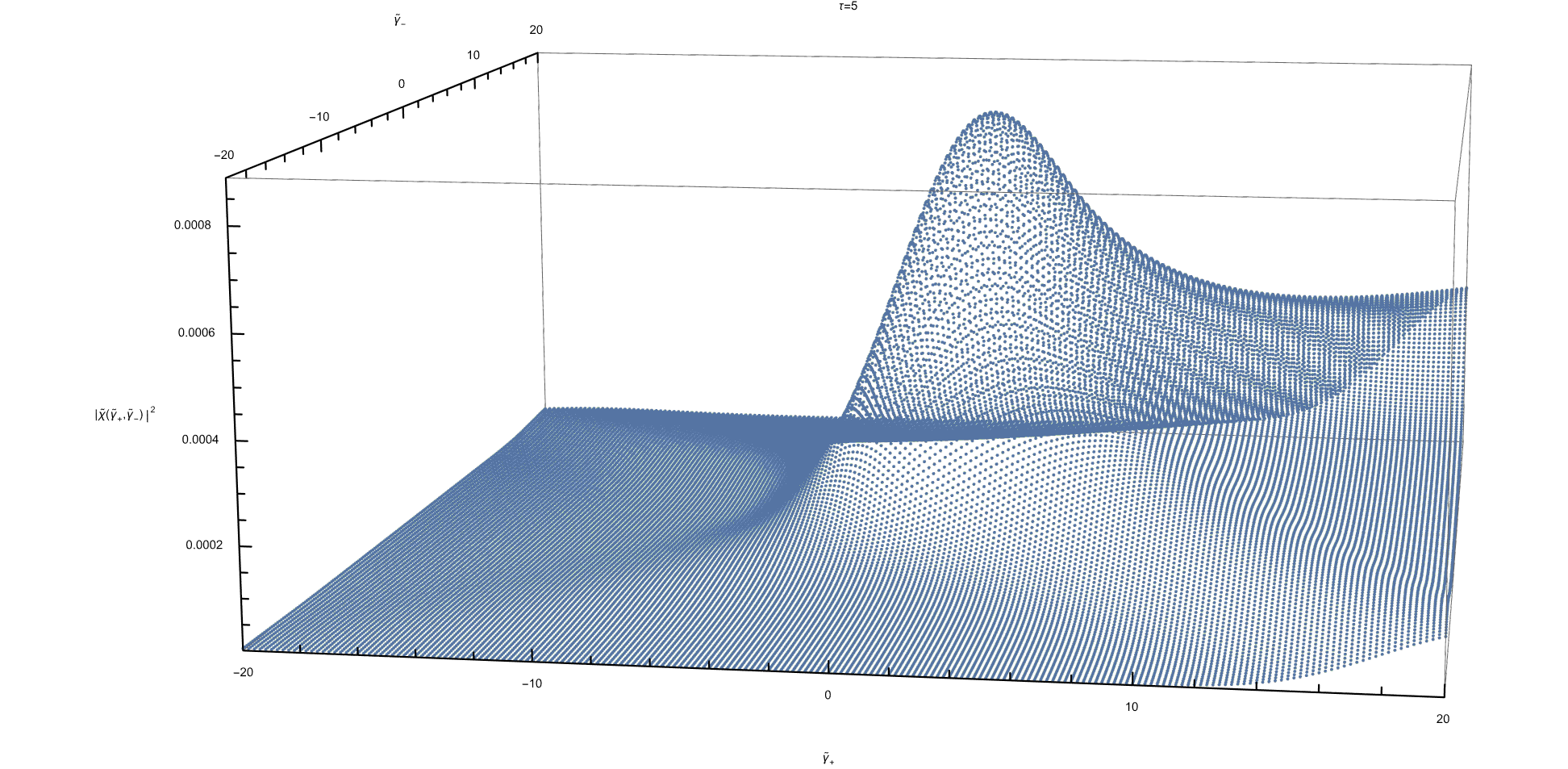}
     \caption{\textbf{Non-commutative case I}:
     dynamics evolution in the time $\tau$ ($\hbar \beta$ units, whit [c]=[$\kappa$]=1) of the Bianchi I wave packet probability density.
     The values set for the parameters of the Gaussian profile are $\nu_{\pm}= 3 $, $\sigma_{\pm}=3.5$, in units of $1/\sqrt{\beta}$.
     It is possible to appreciate the flow of probability mostly along the direction identified by the classical vector of the initial expectation values of momenta and the consequent trajectory of the peak.}
     \label{V_NC_packets}
 \end{figure*}

 \begin{figure*}[!ht]
     \includegraphics[width=%
  	0.325\textwidth]{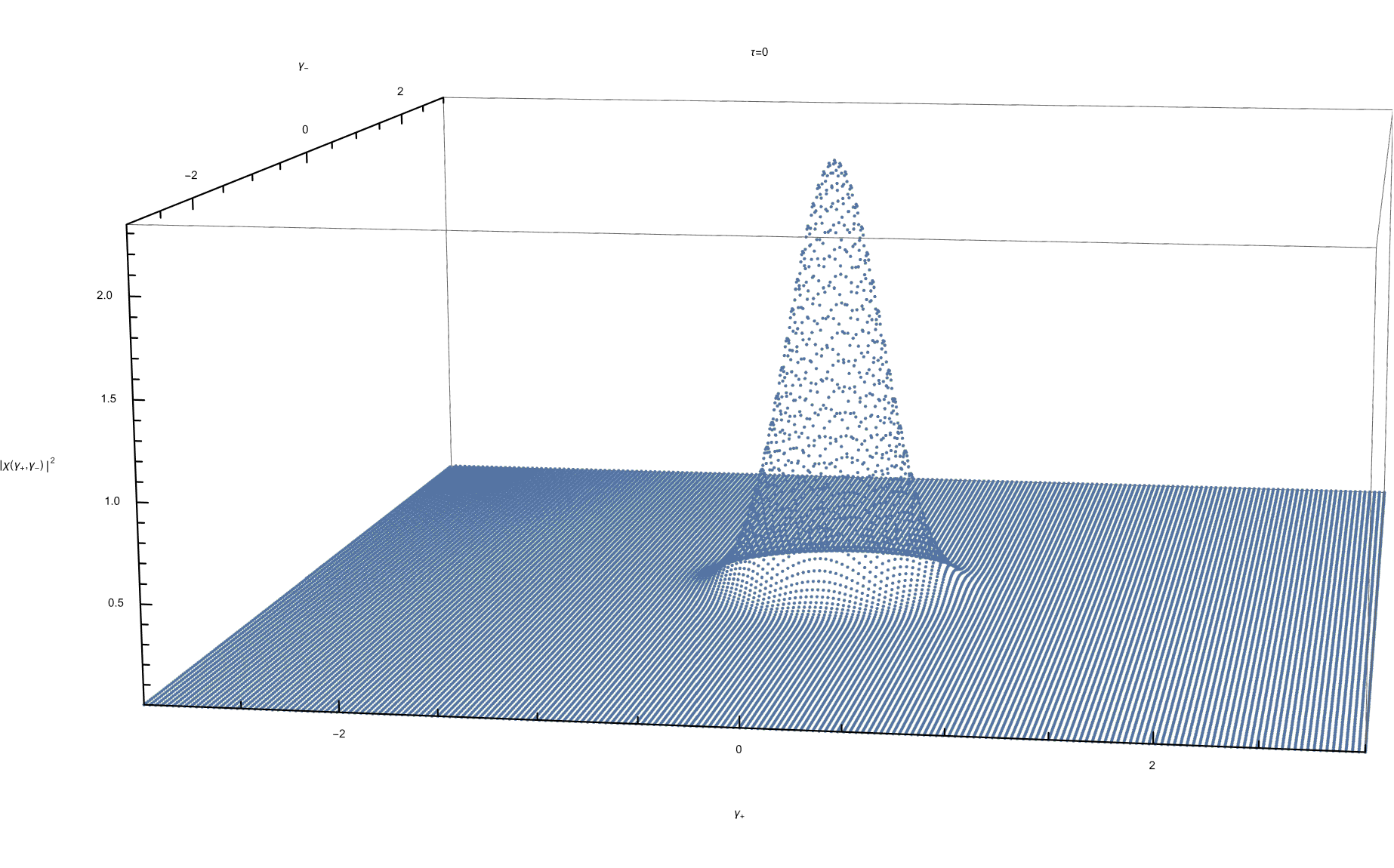}
     \includegraphics[width=%
  	0.325\textwidth]{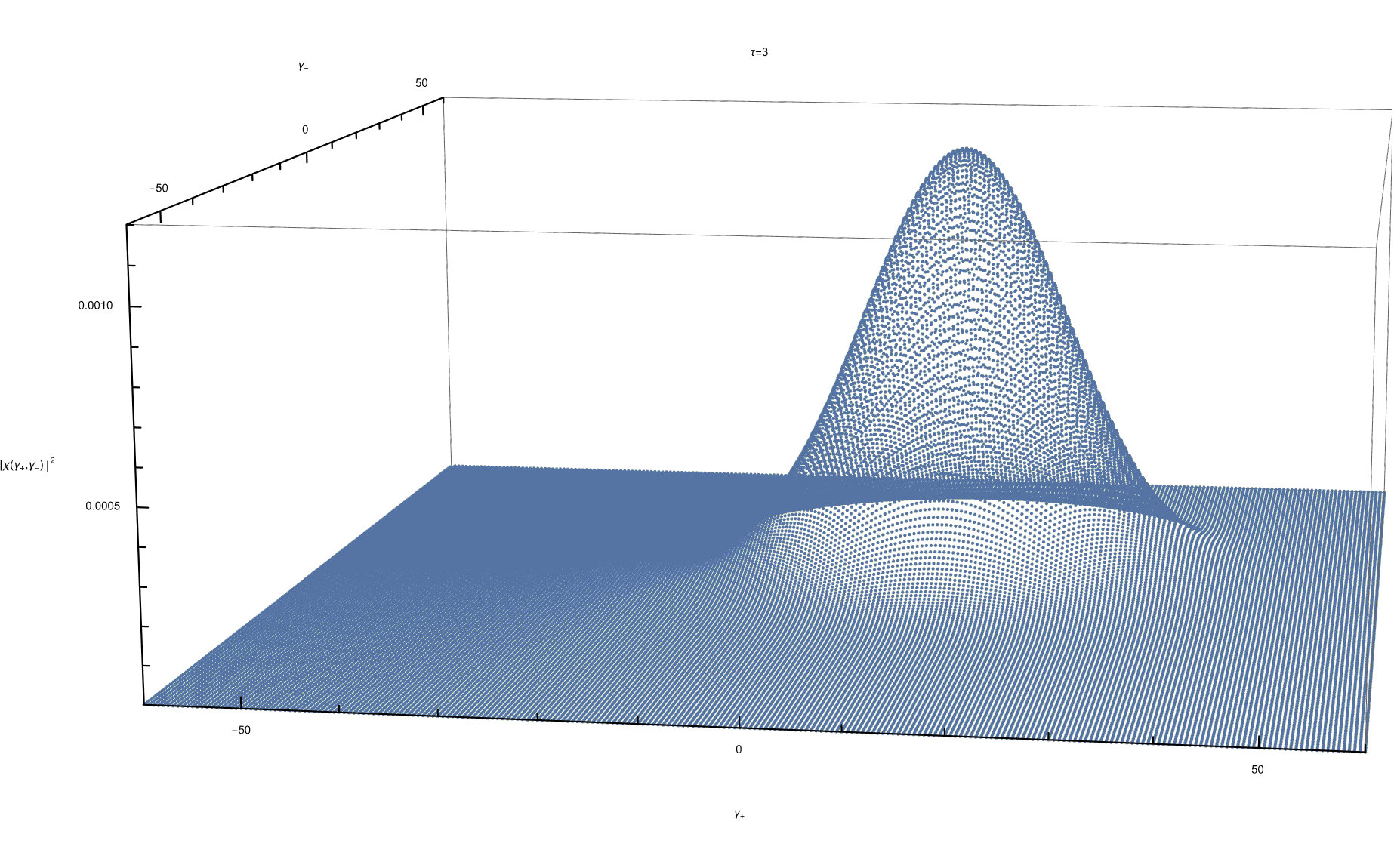}
     \includegraphics[width=% 
     0.325\textwidth]{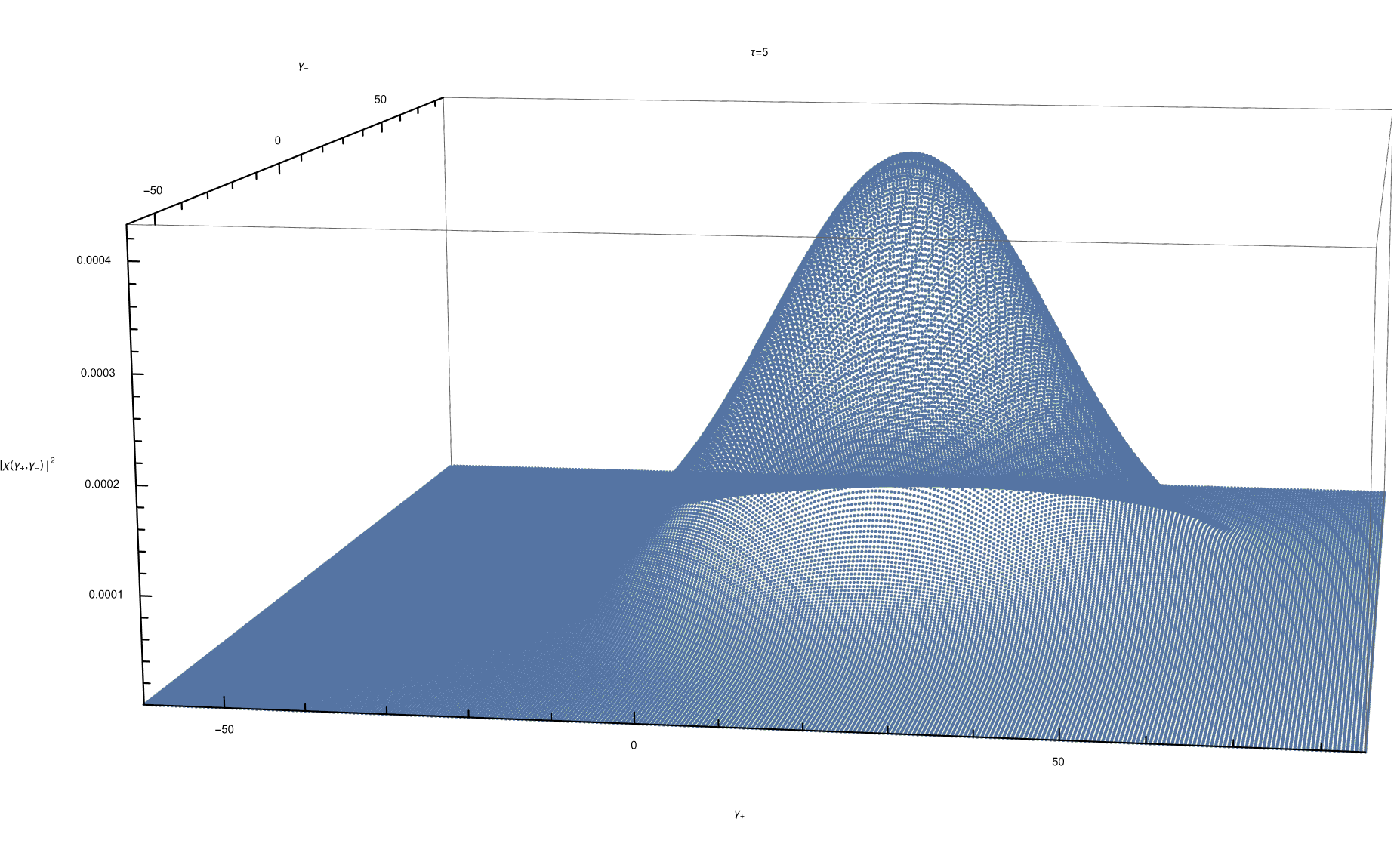}
     \caption{\textbf{Ordinary case I}: dynamics evolution in the time $\tau$ ($\hbar \beta$ units, whit [c]=[$\kappa$]=1) measured in units of $\hbar \beta$, of the Bianchi I wave packet probability density.
     The values of the parameters of the Gaussian profile are fixed in order to match the initial conditions of the corresponding non-commutative wave packet.
     By comparison with the non commutative scenario it is possible to observe the different dynamical behavior of the wave packets, concerning both the spreading geometry and the motion's velocity of the peaks.
     Be aware of the different range of the quasi-anisotropy variables represented in this figure and in Fig~\eqref{V_NC_packets}.}
     \label{V_standard_packets}
 \end{figure*}
 
 As a first thing, we choose a Gaussian profile with a zero position expectation value in the non-commutative variables $(\gamma_{+}, \gamma_{-})$, at the initial time $\tau_0=0$:
  \begin{equation} \label{V_first_gauss_prof}
      \phi_{0}(p_{+},p_{-})\!=\!\exp{\!\!-\!\!\left[ \frac{(p_{+}-\nu_{+})^2}{2 \sigma_{+}^2}+\frac{(p_{-}-\nu_{-})^2}{2 \sigma_{-}^2}\right]},
  \end{equation}
 where $\nu_{\pm}, \sigma_{\pm}$ are real parameters related respectively to the momentum expectation value and uncertainty of the state itself.

 It is important to keep in mind that, in a non-commutative space, we cannot really talk about points, therefore when we refer to a point in the $(\tilde{\gamma}_{+},\tilde{\gamma}_{-})$-space we are actually talking about a region around that point in the not-accessible space $(\gamma_{+},\gamma_{-})$, in the spirit of the quasi-position representation we have constructed.
 In particular, from our results in Sec. \ref{sec_IV}, in this theory the origin is the point in which the localization properties are maximized, hence the extension of this region of uncertainty is the minimal one we can reach.
 From a physical point of view, this means that the origin in the configuration space $(\tilde{\gamma}_{+},\tilde{\gamma}_{-})$ represents the best realization of an isotropic Universe.

\begin{figure*}[!ht]
     \includegraphics[width=%
  	0.32\textwidth]{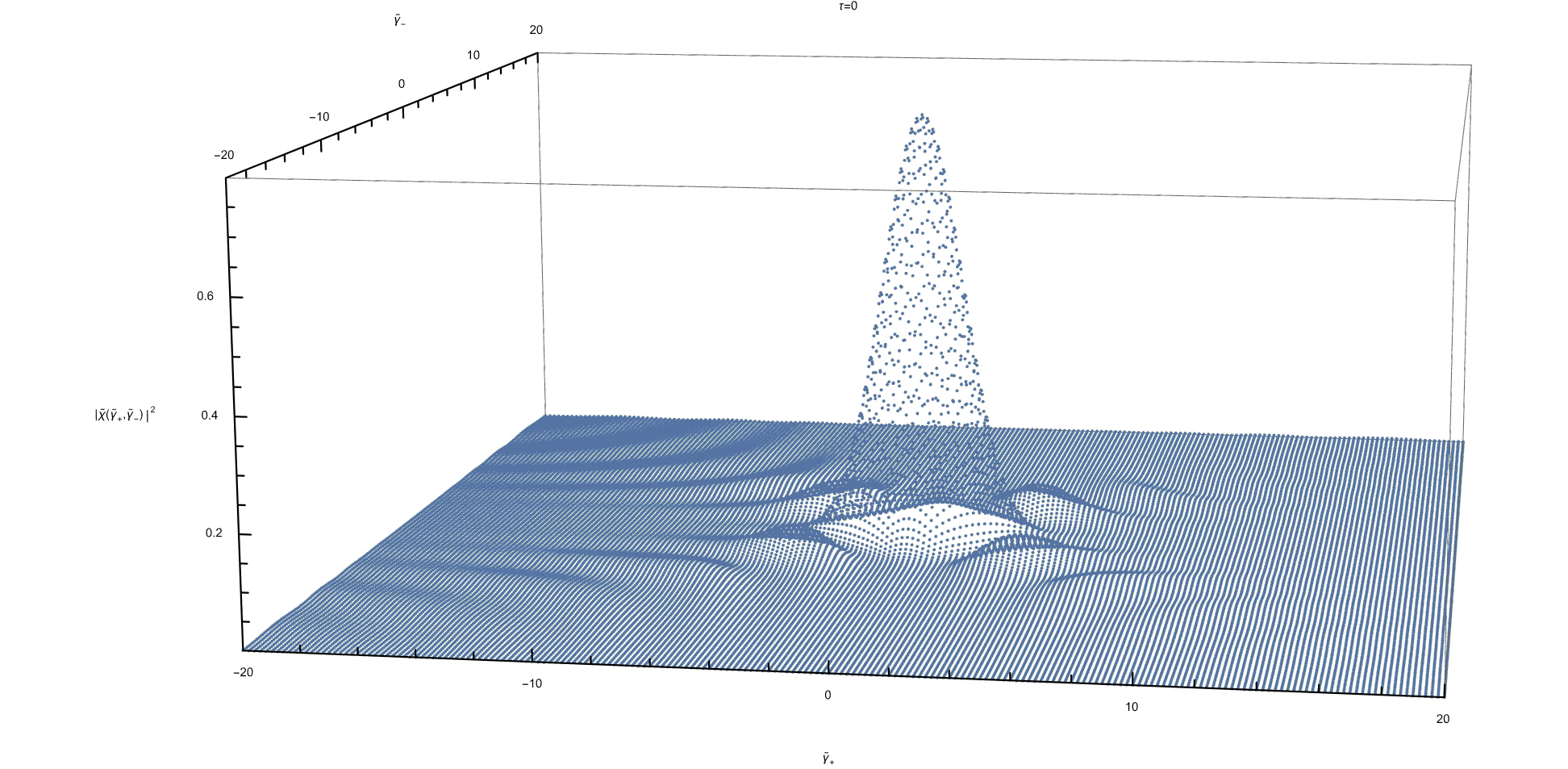}
     \includegraphics[width=%
  	0.32\textwidth]{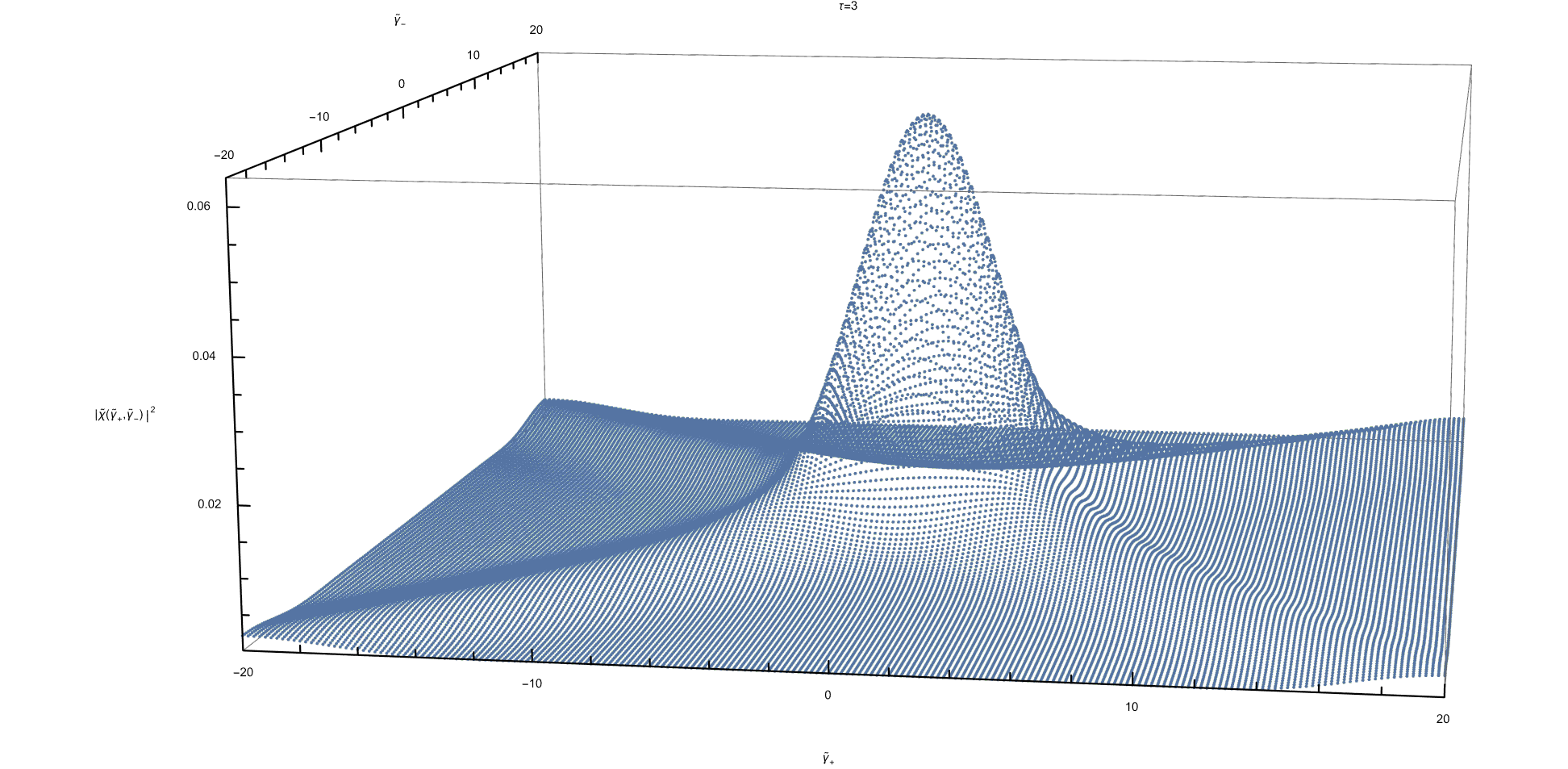}
     \includegraphics[width=% 
     0.32\textwidth]{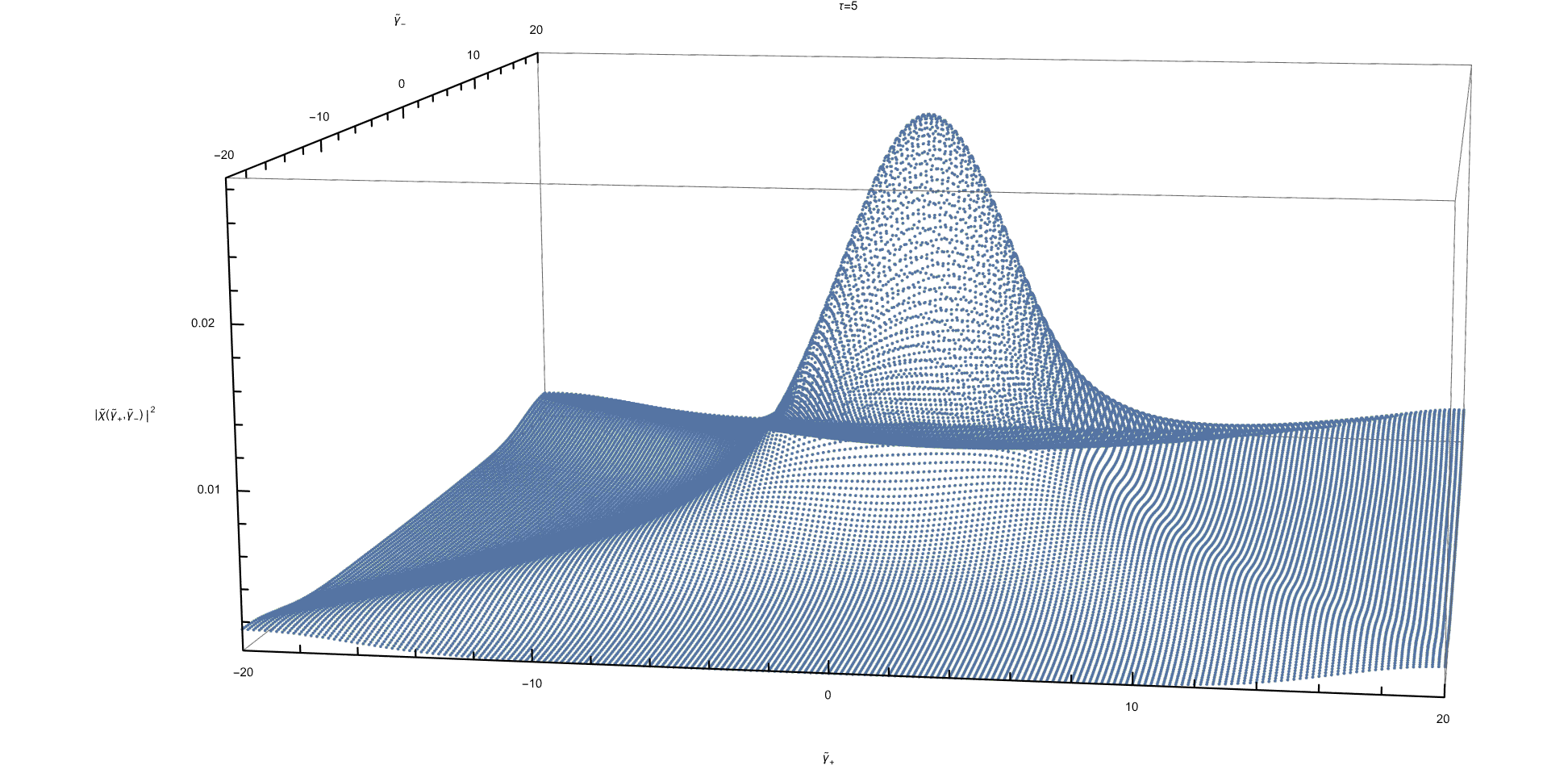}
     \caption{\textbf{Non-commutative case II}:
     dynamics evolution in the time $\tau$ ($\hbar \beta$ units, whit [c]=[$\kappa$]=1) of the Bianchi I wave packet probability density.
     The values set for the parameters of the Gaussian profile are $\nu_{+}= 3.5$, $\nu_{-}=2.5$, $ \sigma_{+}=5.5$, $ \sigma_{-}=4.5 $, in units of $1/\sqrt{\beta}$.
     It is possible to appreciate the \textit{almost} symmetric spread of the wave packet with respect to the origin, in the direction identified by the diagonals of the plane, while it is harder to observe a trajectory of the peak.}
     \label{V_NC_packets2}
 \end{figure*}

 \begin{figure*}[!ht]
     \centering
     \includegraphics[width=%
  	0.32\textwidth]{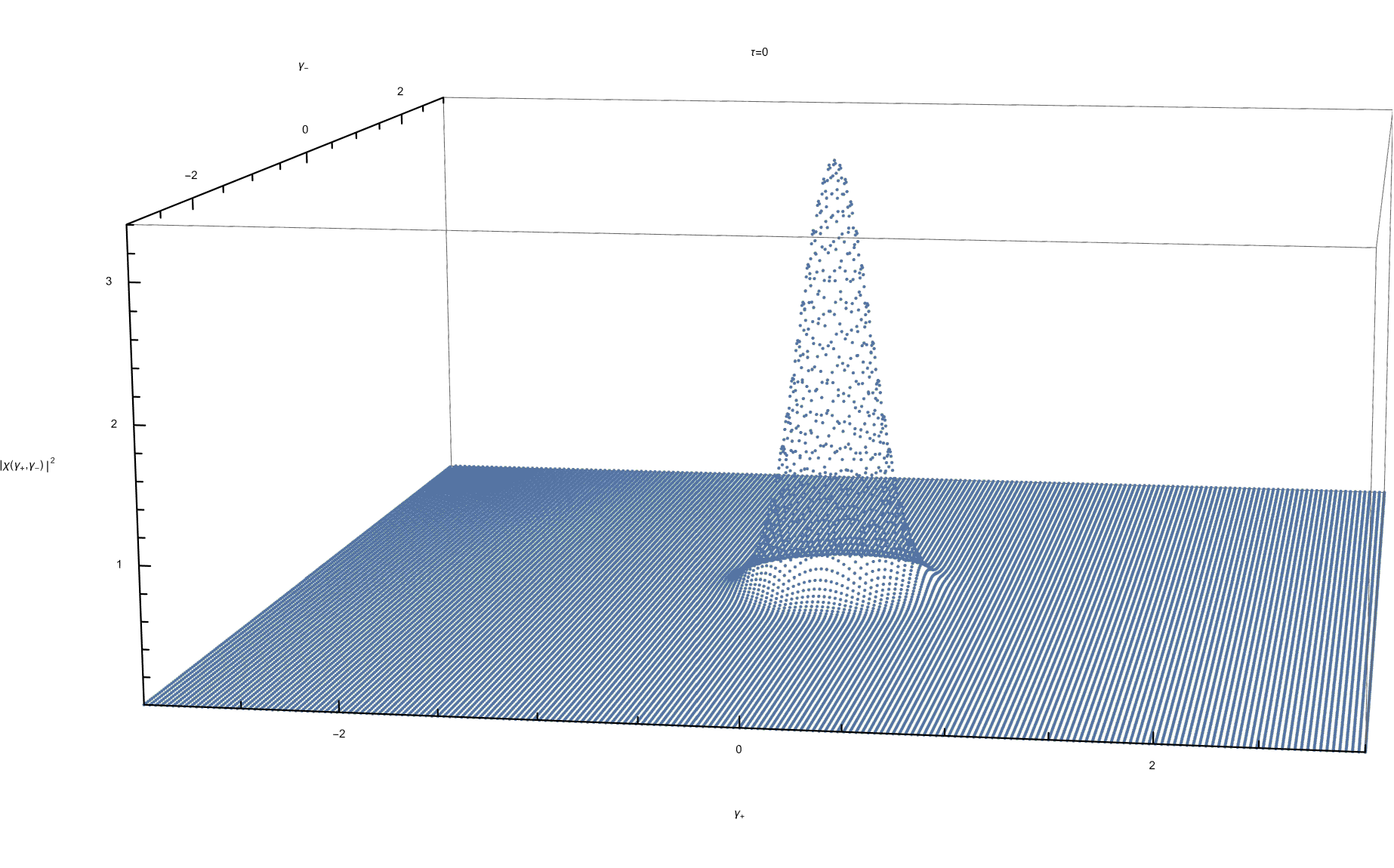}
     \includegraphics[width=%
  	0.32\textwidth]{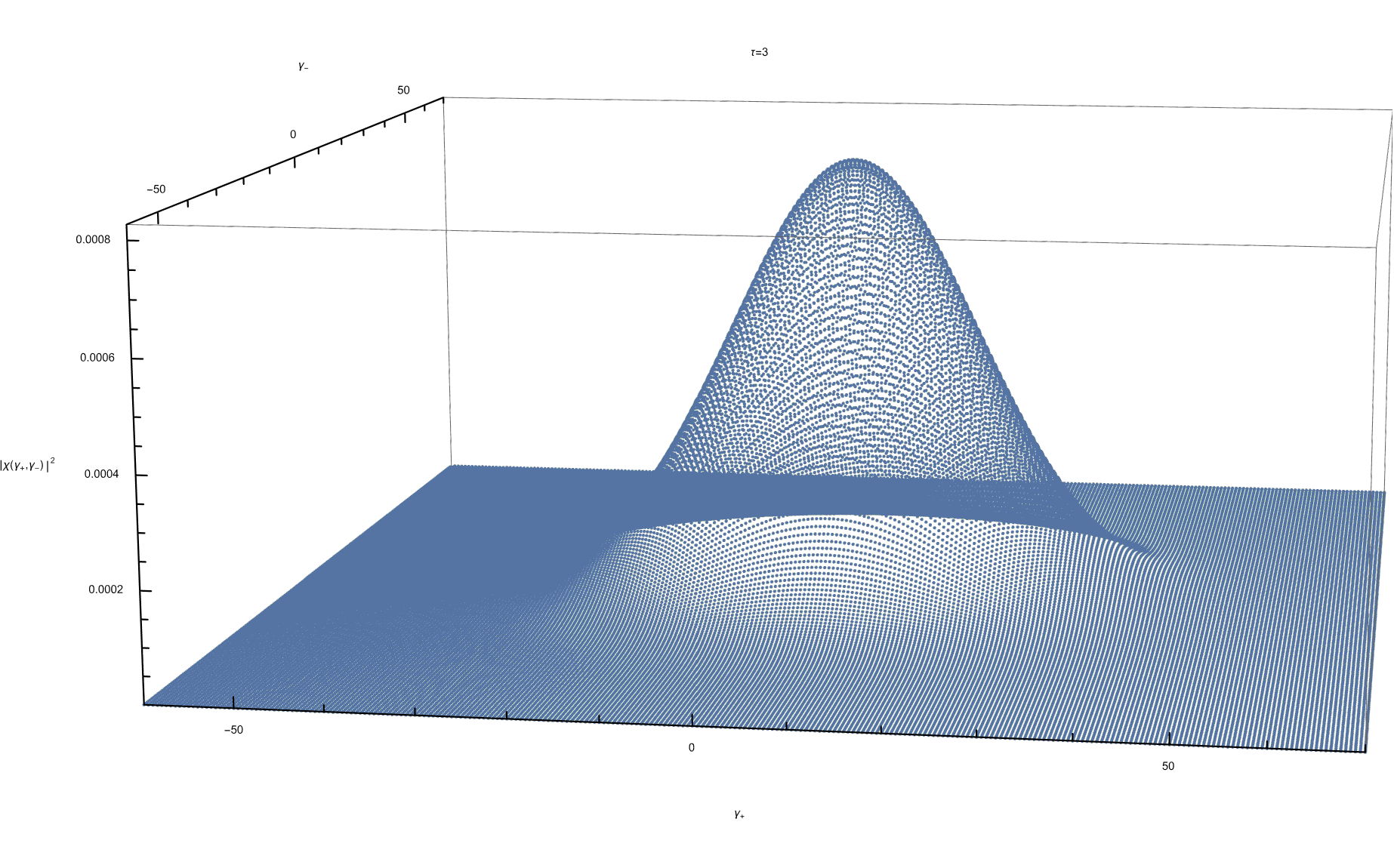}
     \includegraphics[width=% 
     0.32\textwidth]{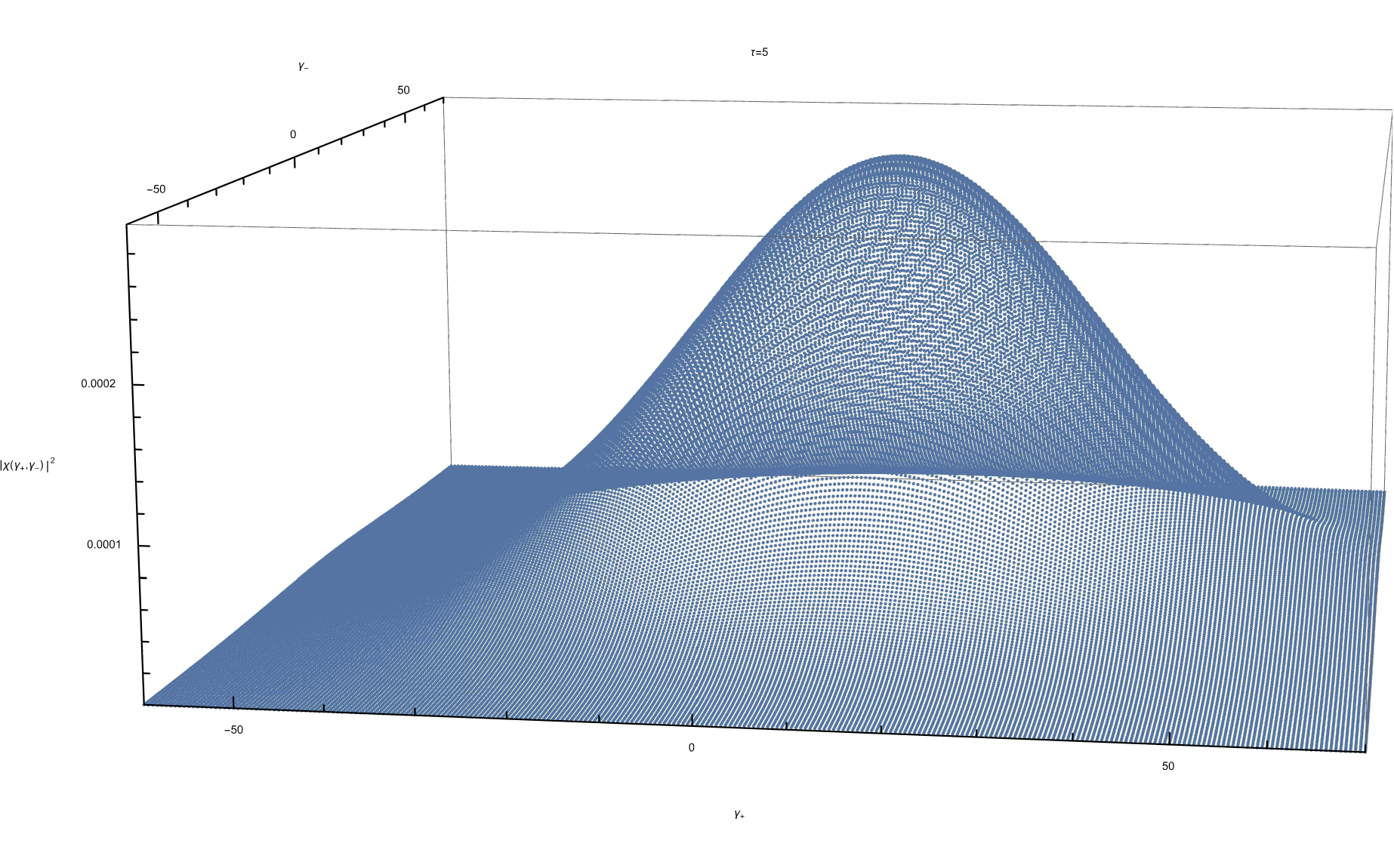}
     \caption{
     \textbf{Ordinary case II}: dynamics evolution in the time $\tau$ ($\hbar \beta$ units, whit [c]=[$\kappa$]=1), of the Bianchi I wave packet probability density.
     The values of the parameters of the Gaussian profile are fixed in order to match the initial conditions of the corresponding non-commutative wave packet.
     By comparison with the non commutative case it is possible to notice the different the dynamical behavior of the wave packets, regarding both the spreading geometry and the motion's velocity of the peaks.
     Be aware of the different range of the quasi-anisotropy variables represented in this figure and in Fig~\eqref{V_NC_packets2}.}
     \label{V_standard_packets2}
 \end{figure*}
 
 Once these parameters are properly fixed, numerical integration of \eqref{V_general_sol_SE} with the profile \eqref{V_first_gauss_prof} allows us to follow the evolution of the quantum Universe at different times $\tau$.
 Two different choices for set of the involved parameters has been done, in order to distinguish between two general cases, as shown in Fig~\eqref{V_NC_packets} and Fig~\eqref{V_NC_packets2}.
 In the first scenario, in the profile \eqref{V_first_gauss_prof}, we have fixed $\nu\_{\pm} > \sigma_{\pm}$, while in the second we have set  $\nu\_{\pm} < \sigma_{\pm}$.

 By visual inspection, we can infer features of our non-commutative quantum model and appreciate the difference between the two general cases.
 In both scenarios, while time elapses, the system is spreading, signaling that the probability of finding the particle Universe in another region of the configuration space $(\tilde{\gamma}_{+},\tilde{\gamma}_{-})$ away from the origin increases in time.
 Nevertheless, the geometry of the two spreading processes is quite different.
 While in the first case the probability density of the wave packet is flowing \textit{mostly} along a precise direction, which is the one \textit{roughly} identified by the classical vector of the initial expectation values of momenta, in the second case the wave packet is spreading \textit{almost} symmetrically with respect to the origin, in the direction identified by the diagonals of the plane.
 
 \noindent 
 
 The peak of the wave packets will move somehow accordingly:
 in the scenario depicted in Fig~\eqref{V_NC_packets}, it follows the flow of probability density and it is possible to appreciate its trajectory; in the scenario presented in Fig~\eqref{V_NC_packets2}, due to the different spreading progress, the peak's movement is much slower and, consequently, it is harder to observe a trajectory.
 
 Finally we note the different initial localization properties of the two wave packets in the quasi-anisotropies space, namely in the second example the wave packet appears to be more localized with respect to the first one.

 This behavior of the Universe wave packet in this non-commutative setting is remarkably different with respect to the behavior of a properly defined wave packet describing the same Universe, namely Bianchi I, in the ordinary quantum theory.

 As it is clear from Fig~\eqref{V_standard_packets} and Fig~\eqref{V_standard_packets2} and the visual comparison with the respective non-commutative counterparts, the ordinary Bianchi I wave packet does not display any different spreading geometries in the two illustrations and it moves from the initial point towards regions with higher absolute values of the anisotropies, with a greater velocity with respect to the non-commutative cases.

 Due to this tendency, the initial configuration loses "rapidly" and in a definitive way the status of favored one, differently from what happens in the non-commutative framework, where this point remains the one in which we can find the particle Universe with a higher probability for a longer time if compared to the ordinary quantum dynamics.

 \noindent

 To make sense of this qualitative statement, considering as initial configuration the isotropic one, we can compare how the relative probability density at the origin of the wave packets Fig~\eqref{V_NC_packets} and Fig.\eqref{V_standard_packets} changes over time.
 
  \noindent 
  
  The situation for the wave packets of the second scenario is identical.

\begin{figure}[!ht]
     \includegraphics[width=%
  	0.48\textwidth]{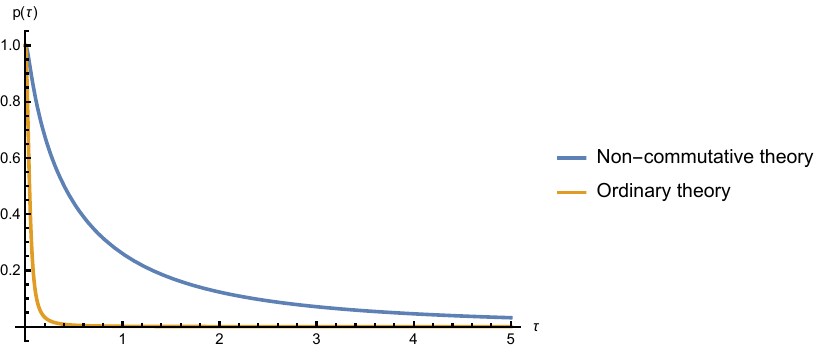}
     \caption{Plot of the relative probability density $p(\tau)=\abs{\Psi(0,0,\tau)}^2/\abs{\Psi(0,0,0)}^2$ at the origin concerning the non-commutative wave packet of Fig~\eqref{V_NC_packets} and the ordinary wave packet of Fig~\eqref{V_standard_packets}, where $\Psi$ stands for the generic wave function.
     It can be noticed how the relative probability in the non-commutative case prevails over the relative probability of the ordinary theory at any displayed time $\tau$.}
     \label{V_rel_prob1}
 \end{figure}

 In Fig~\eqref{V_rel_prob1}, it can be appreciated how the point $(0,0)$, denoting the isotropic state in both theories, experiences a faster relative decline in probability within the ordinary theory compared to the non-commutative one. This suggests that in the non-commutative theory, the isotropic configuration (or any initial configuration, actually) maintains a greater probabilistic advantage with respect to the ordinary theory.

 The same scenario is verified also if we initially peak our wave packet in an anisotropic configuration: the particular point chosen for the initial condition will be the favored one for a relatively long time, again if compared to what happens to the ordinary wave packet.

 Here it is important to stress that in our theory it only makes sense to explore "small" regions around the origin. This is true for two orders of reasons:
 \begin{itemize}
     \item the Vilenkin approximation we have used constrains us to remain in a "small" quantum phase space (see above);
     \item in our non-commutative theory we can talk about localization only in proximity of the origin.
 \end{itemize}

 These insights emphasize how the non-commutative GUP quantum theory modifies some aspects of the dynamics of the Bianchi I Universe, establishing new different relationships between the initial probability densities and their evolution, concerning both isotropic and anisotropic configurations of the Universe.
 
 \section{Conclusions} \label{sec_VI}

  In our analysis, we faced a subtle and challenging problem, concerning the
  construction of a consistent quantum theory for a $n$-dimensional non-commutative GUP formulation of the Heisenberg algebra.
  Our main concern specifically regarded the possibility of recovering a way to describe the localization properties of the theory.
  After a brief extension of the necessary functional analysis of the fundamental operators, as a first result, we were able to identify those states that minimize the uncertainty in the coordinate operator along a specific direction.
  The price to pay in order to achieve this maximal localization in one of the space coordinates is to delocalize the state itself of a certain amount in all other directions.
  This means that, in general, it is not possible to achieve at the same time a maximal localization with respect to all the $n$ variables.
  This fact is a natural manifestation of the non-commutative nature of the algebra we are considering.

  Since these states alone cannot be used to describe the position on the whole configuration space, we defined a procedure to construct a consistent generalization of the quasi-position representation introduced in \cite{Kempf:1994su}.
  We successfully identified the only state of the theory that exhibits isotropic properties of localization, that is that state which admits simultaneously the same maximal localization in all the $n$ variables.
  This state is centered around the origin of the system of coordinates and it is a spherically symmetric state in momentum space.
  The translation - properly defined with respect to the expectation position value - of this state in all directions of a certain arbitrary amount, allowed us to generate a family of states suitable to be used as a quasi-position basis.

  Once reached the task of setting up a consistent quantum theory for the non-commutative $n$-dimensional GUP, we provided a two-dimensional implementation of our quantum scenario to the evolution of the Bianchi I cosmological model in the WKB approximation of a quasi-classical volume, while dealing with pure quantum anisotropy degrees of freedom.

  More precisely, by adopting the standard Misner variables, we identified the so-called isotropic variable $\alpha$ as the one approaching a quasi-classical limit and the two variables $\gamma_{\pm}$ with the ones belonging to a small quantum subsystem.
  In cosmology, the two anisotropy degrees of freedom play the role of the two physical degrees of the gravitational field, while $\alpha$ is often called an "embedding'' variable and, here, it restores, via its classical limit, the dependence of the anisotropy variables on the label time $t$.
  Accordingly, in a scenario à la Vilenkin, we were able to arrive at a real
  two-dimensional Schr\"odinger equation in momentum space for the description of the dynamics of the quantum part of the Universe, now dictated by our GUP-modified Heisenberg algebra.

  By analyzing wave packets describing our Bianchi I Universe, we were able to show how differently these wave functions behave with respect to the ordinary quantum case.
  
  We showed how these wave packets exhibit more complex spreading properties, with a "geometry" depending on the their initialization in momentum space, and we observed how their overall motion in the quasi-anisotropies plane - identified with the motion of the peak - is slower in time when compared to the ordinary quantum theory.
  
  This particular tendency makes the initial configurations more "stable", in the sense that, from a probabilistic point of view, they are the favored ones for a time which is longer if compared with the same situation in the ordinary theory.

  This holds for any initial configuration in the permitted region restricted by the Vilenkin approximation  (see \cite{Agostini:2017oql}) and the localization properties of the GUP framework and clearly also for an isotropic one, that is a configuration in which the wave packet describing our Universe is peaked and sufficiently localized around the origin of the plane $(\tilde{\gamma}_+,\tilde{\gamma}_-)$.
  
  \noindent
  The observed behavior suggests the idea that non-trivial Heisenberg algebras, thought in the Mini-superspace as the way for accounting of cut-off physics, can potentially provide new physical insights on some open questions of quantum cosmology, such as the \textit{classicalization} of the Universe \cite{Kirillov:1997fx}, the emergence of an isotropic Universe from more general configuration, which appear to be very natural in the proximity of the initial singularity \cite{Imponente:2001fy, Benini:2006xu} or the possibility to generate Big Bounce scenarios, as in the case of Polymer quantum mechanics \cite{Barca:2021qdn} and Loop quantum cosmology \cite{Ashtekar:2009vc, Bruno:2023aco, Bruno:2023all}. 
  
  More complex Bianchi models with richer structures will be for sure the best candidates to test the validity of these ideas and hence the effectiveness of this non-commutative GUP quantization framework.

 \appendix 
 \section{Computation of the coordinate operator uncertainty in the i-th direction} \label{A_1}
 
     The computation of the uncertainty in the $i$-th coordinate for the state \eqref{IV_quasi_pos_basis} essentially relies on the computation of the integral quantity:
      \begin{align}
          \bra{\Phi_{T}^{ml}}{\hat{\mathbf{x}}_{i}^{2}} \ket{\Phi_{T}^{ml}}=- \!\!\! \int_{\mathbb{R}^n} & d^n p \left(\Phi^{ml}_{T}(p)\right)^{*} \left[2 p_{i} \partial{p_{i}}\Phi^{ml}_{T}(p) \right.  \nonumber \\ 
          & \left. + (1+ p_{r}p_{s}\delta^{rs})\partial^{2}_{p_{i}}\Phi^{ml}_{T}(p)\right] \frac{\hbar^2 \beta}{\beta^{n/2}},
      \end{align}
    where $i,j$ run over all the $n$ indices.
    Notice that we have factored out the dimensional quantities in order to work, inside the integral, with adimensional objects.
    
    Every resulting integrals can be resolved by using hyperspherical coordinates.
    
    The first integral, i.e. the one involving the first partial derivative, once made explicit, has only one contributing term:
     \begin{align}
         & \int_{\mathbb{R}^{n-1}}\!\!\!\!  d^{n-1 }  p \int_{\mathbb{R}} dp_{i} \, 2 p_{i} (1+ p_{r}p_{s}\delta^{rs})^{-n}\left[\frac{-n p_i + I \xi_{i}}{1+p_r p_s\delta^{rs}}\right]\nonumber \\
         & =\int d\Omega_{n-1}\int_{\mathbb{R}^{+}} dr \; r^{n-2}(1+r^2)^{1/2-n}\sqrt{\pi}\frac{\Gamma[n-\frac{1}{2}]}{\Gamma[n]} \nonumber \\
         & =\frac{2^{1-n}\pi^{\frac{n+1}{2}}}{\Gamma[\frac{n+1}{2}]},
     \end{align}
     where $r:=\sum_{l\neq i} p_{l}^2$ and $\Omega_{n-1}$ is the solid angle in $(n-1)$ dimensions.
     
    The second integral, i.e. the one involving the second partial derivative, can be decomposed in different contributing terms.
    
    The first term is given by:
     \begin{align}
         &\int_{\mathbb{R}^n}d^n p \left[-n(2+n)p^2_{i}(1+p_r p_s \delta^{rs})^{-1-n}\right] \nonumber \\
         &=\int_{\mathbb{R}^{n-1}} \!\!\!\!\! d^{n-1}p\int_{\mathbb{R}} dp_{i} \left[-n(2+n)p^2_{i}(1+p_r p_s \delta^{rs})^{-1-n}\right] \nonumber\\
         &=\!\! \int \!\!\!\, {d \Omega_{n-1}} \!\! \int_{\mathbb{R}^{+}} \!\!\!\!\!\! dr \, r^{n-2} \!\! \left( \!\! \frac{-n(2+n)\sqrt{\pi}(1+r^2)^{\frac{1}{2}-n} \Gamma\left[n-\frac{1}{2}\right]}{2 \Gamma\left[n+1\right]}\! \right) \!  \nonumber\\
         &= - \frac{2^{-n}(2+n)\pi^{\frac{1+n}{2}}}{\Gamma\left[\frac{n+1}{2}\right]}, \qquad \text{with}\quad  r:=\sum_{l\neq i} p_{l}^2
    \end{align}
   The second term is:
     \begin{align}
          &\int_{\mathbb{R}^n}\!\!d^n p (1+p_r p_s \delta^{rs})^{-n}= \! \int \!\! d{\Omega_{n}} \int_{\mathbb{R}^{+}}\!\!\!\!\!dr \; r^{n-1} n(1+r^2)^{-n}\!  \nonumber \\
          & = \frac{2^{1-n}n \pi{\frac{1+n}{2}}}{\Gamma\left[\frac{n+1}{2}\right]},  \qquad \text{with}\quad  r:=\sum_{l \neq i} p_{l}^2
     \end{align}

   Finally, the third and last term, which is the most complex one, is the following:
    \begin{align}
         &  \int_{\mathbb{R}^n} \!\!\!\!\!\! d^n p (1+p_r p_s \delta^{rs})^{1-n}\!\! \left\{\!\!\left[\sum_{\substack{k=1 \\ k \neq i}}^{n} \left(\!\frac{\!p_{i}p_{k} \xi_{k}\!}{\left(1+ p_{r}p_{s} \delta^{rs}|_{r,s \neq k}\right)}\!\right) \!- \! \xi_i \!\right] \right. \nonumber \\
         &\left. \times \frac{1}{\left(1+ p_{r}p_{s} \delta^{rs}\right)}\! + p_{i} \!\! \left[ \sum_{\substack{k=1 \\ k \neq i}}^{n}\frac{\xi_{k} \tan[-1](\!\frac{p_{k}}{\sqrt{1+ p_{r}p_{s} \delta^{rs}|_{r,s \neq k}}}\!)}{(1+ p_{r}p_{s} \delta^{rs}|_{r,s \neq k})^{3/2}}\right] \!\! \right\}^{\!2}\!\!\!.
    \end{align}
  We can further identify three pieces in this last integral, one for every term of the square.

  The sum of first two integrals, that is the square of the first term and the mixed term, gives the following results:
   \begin{align}
       \sum_{\substack{k=1 \\ k \neq i}}^{n} & \left[  - \frac{\pi^{n/2}\xi_k^2 \Gamma\left[1 +\frac{n}{2}\right]}{n \Gamma[n]-4n^3 \Gamma[n]} + \frac{n \pi^{n/2}\xi_k^{2} \Gamma[\frac{n}{2}]}{4n^3-4n^2-n+1)\Gamma[n]} \right] \nonumber \\
       & \times \frac{2^{-n}\pi^{\frac{n+1}{2}}\xi^{2}_{i}}{\Gamma[\frac{n+1}{2}]}.
   \end{align}

  The technique to resolve these integrals is once again the same as the previous used above:
  first we integrate in those $p$-variables which do not enter the integral in a rotational invariant way, then, once the integral is reduced to only a $n$-dimensional spherical symmetric term we employ hyperspherical coordinates to compute it.

  The resolution of the last integral, that is the one containing the square of the arctangent term, is more involved.
  
  The integral we need to solve is:
   \begin{align} \label{A_int_q_I}
       & \sum_{\substack{k=1 \\ k \neq i}}^{n} \int_{\mathbb{R}^n} \!\!\! d^n p (1+p_k^2+p_{l}p_{m}\delta^{lm})^{(-n+1)}p_{i}^2 \xi_{k}^2    \nonumber \\
       & \times \left[\tan[-1](\frac{p_k}{\sqrt{1+p_{l}p_{m}\delta^{lm}}})\right]^2 \frac{1}{\left(1+p_k^2+p_{l}p_{m}\delta^{lm}\right)^3},
   \end{align}
   where we remember that $l,m \neq k$.
   
   Considering just a generic term of the sum, we make a change of variable, namely $q:=\arctan[\frac{p_{k}}{\sqrt{1+p_i^2 + r^2}}]$, where we have set $r:=p_{l}p_{m}\delta^{lm}|_{l,m \neq k,i}$.
  The integral \eqref{A_int_q_I} then become:
   \begin{equation}
       \xi_{s} ^2 \! \int_{\mathbb{R}^{n-2}} \!\!\!\!\!\! d^{n-2}p \int_{\mathbb{R}} dp_{i} \,  p_{i}^2 (1+p_i^2 + r^2)^{-\frac{3}{2}-n}\!\! \int_{-\pi/2}^{\pi/2} \!\! dq   \frac {q^2}{\cos[{4-2n}](q)}
   \end{equation}
  The $q-$integral can be solved for every $n>1$ but it does not admit a closed form.
  
  By induction we can show that the general result can be written as:
   \begin{equation}
       \int_{-\pi/2}^{\pi/2} dq   \frac {q^2}{\cos[{4-2n}](q)} = \alpha(n) \pi + \theta(n) \pi^3,
   \end{equation}
  where $\alpha(n)$ is a real negative function and $\theta(n)$ is a real positive function, both defined for $n>1$.
  A table with the value of the integral for some values of $n>1$ can be find below \eqref{A1_table1}.
  \begin{table}
  \centering
  \begin{tabular}{||l||c|r||}
   \hline
    $n$ & $\alpha(n)$ & $\theta(n)$ \\
   \hline
   \hline
    2 & 0 & $\frac{1}{12}$\\
   \hline
    3 & $-\frac{1}{4}$ & $\frac{1}{24}$\\
    \hline
    4 & $-\frac{15}{64}$ & $\frac{1}{32}$\\
    \hline
    5 & $-\frac{245}{1152}$ & $\frac{5}{192}$\\
    \hline 
  \end{tabular}
  \label{A1_table1}
  \caption{Table with the values of the real functions $\alpha(n)$ and $\theta(n)$ for some $n$, where $n \in \mathbb{N}$ is the number of dimensions in which we are working.}
  \end{table}

  Considering this, by performing the integration in $p_{k}$ and then using again hyperspherical coordinates for the last integration, in the end we obtain:
     \begin{equation}
        \left(\alpha(n) \pi +\theta(n) \pi^3\right) \pi^{\frac{n-1}{2}}\frac{\Gamma[1+\frac{n}{2}]}{2 \Gamma[\frac{3}{2}+n]}  \left[ \sum_{s \neq k} \xi_{s}^2\right]
     \end{equation}
  where we have re-introduced the initial sum over all $s \neq k$.

  Finally - summing all the contributes properly re-scaled with the norm of the state shown in \eqref{IV_quasi_pos_basis} and considering the dimensional quantities factored out at the beginning - after some algebra involving properties of the Euler Gamma function, we are able to obtain the expression \eqref{IV_unc_pos_nc_basis}, concerning the uncertainty of the coordinate operator in a generic direction $k$, for the state \eqref{IV_quasi_pos_basis}:
     \begin{align}
          (\Delta{\hat{\mathbf{x}}_i})^2=\frac{1}{2}&\left\{ 2n+\frac{2+4n}{(n-1)(4n^2-1)}\sum_{k\neq i} \xi^2_{k}  \right. \nonumber \\
        &\left. \!\!\!\! + \frac{n!}{\Gamma[3/2+n]}\sqrt{\pi}\biggl(\alpha(n)+\theta(n)\pi^2\biggr) \sum_{k\neq i} \xi^2_{k} \right\} \hbar^2 \beta.
     \end{align}

  \section{Inverse transform} \label{A_2}

   In Section \ref{sec_IV} we have defined the integral transform \eqref{IV_map_mom_to_quasipos}, which maps wave functions from momentum space to quasi-position space:
    \begin{align}
        \psi(\xi):=  & \int_{\mathbb{R}^n} \frac{d^n p}{(1+ \beta p_rp_s \delta^{rs})}\prod_{k=1}^{n} e^{i \xi_{k}g\left(p_k, p_{\{l\}\neq k}\right)} \nonumber \\
        & \times (1+ \beta p_rp_s \delta^{ij})^{-n/2}\psi(p),
    \end{align}
    where we have defined:
     \begin{equation}
         g\left(p_k, p_{\{l\}\neq k}\right):=\frac{\tan[-1](\frac{\sqrt{\beta}p_{k}}{1+\beta\sum_{l}p^2_{l}})}{\sqrt{1+\beta\sum_{l}p^2_{l}}}, \qquad \forall k
     \end{equation}
    and $i,j$ run over all the $n$ indices.
    
   To construct the inverse of this transform we can rewrite the previous relation as:
    \begin{align} \label{B_inv_int_1}
       &\int_{\mathbb{R}^n} d^{n} \xi \prod_{k=1}^{n} e^{-i \xi_{k} g\left(p'_k, p'_{\{l\}\neq k}\right)} \psi(\xi) \nonumber \\
       & = \int_{\mathbb{R}^n}  \prod_{k=1}^{n} e^{-i \xi_{k} g\left(p'_k, p'_{\{l\}\neq k}\right)} \left(\int_{\mathbb{R}^n} \frac{d^n p}{(1+ \beta p_rp_s \delta^{rs})} \right. \nonumber \\
       & \left. \times \prod_{k=1}^{n} e^{i \xi_{k} g\left(p_k, p_{\{l\}\neq k}\right)}(1+ \beta p_ip_j \delta^{ij})^{-n/2}\psi(p)\right) d^{n} \xi.
   \end{align}
  By exchanging the order of integration and performing the $\xi$-integral, we can write for the right-hand side of \eqref{B_inv_int_1}:
   \begin{align} \label{B_inv_int_2}
       (\sqrt{2 \pi})^{n}\int_{\mathbb{R}^n}& d^n p \, (1+ \beta p_r p_s \delta^{rs})^{-\frac{n}{2}-1}\psi(p) \\
       &\times \prod_{k=1}^{n} \delta\left(g\left(p_k, p_{\{l\}\neq k}\right)-g\left(p'_k, p'_{\{l\}\neq k}\right)\right).
   \end{align}
 The presence of the deltas inside the integral define a specific region of $\mathbb{R}^n$ on which the integral itself has to be evaluated. 
 This region is given by the intersection of $n$ $(n-1)$-dimensional hypersurfaces in a $n$-dimensional space, hence corresponds to a point or a set of points.
 
 To explicitly perform the integration we need to use the so-called \textit{coarea formula} \cite{Federer1969GeometricMT}.

 Let us consider the integral over an open set $A$ of $\mathbb{R}^n$ of an integrable function $f(x)$.
 
 Introducing the vector function $u(x)$:
  \begin{equation}
      u: A \subseteq \mathbb{R}^n \to \mathbb{R}^k, \qquad k\leq n,
  \end{equation}
 the coarea formula states that:
  \begin{equation} \label{B_coarea}
      \int_{A} f(x) d^{n}x:= \int_{\mathbb{R}^k}\left(\int_{u^{-1}(t)} \frac{f(x)}{\abs{\mathcal{J}_k u(x)}} d H_{n-k}(x)\right) d^{k} t
  \end{equation}
 where $\abs{\mathcal{J}_k u(x)}$ is the determinant of the Jacobin matrix of $u(x)$, which can be defined also in the rectangular case, and the measure in the second integral is the Hausdorff measure in $(n-k)$ dimensions.

 We want to apply this formula to our integral \eqref{B_inv_int_2}. 
 
 Rigorously speaking, due to the Dirac deltas, our function is not an integrable one over $\mathbb{R}^n$.
 Nevertheless we can define the delta as the limit of a suitable integrable function in order to deal with a proper overall integrable function in every steps and, in the end, consider the limit itself to reintroduce the deltas.
 In this way we consider our integral containing delta distributions well-defined with respect to the possibility of applying the theorem.

 In order to proceed, first, we define two vector fields, namely:
  \begin{align}
     &\qquad \qquad \qquad G: \mathbb{R}^n \to \mathbb{R}^n  \nonumber \\
     &G(p):=\left(g\left(p_1, p_{\{l\}\neq 1}\right),...,g\left(p_n, p_{\{l\}\neq n}\right)\right)
  \end{align}
 and
 \begin{align}
     &\qquad \qquad \qquad V: \mathbb{R}^n \to \mathbb{R}^n  \nonumber \\
     & V(p):=\left(g\left(p_1, p_{\{l\}\neq 1}\right)-g\left(p'_1, p'_{\{l\}\neq 1}\right),..., \right. \nonumber \\
     &\left. \qquad \quad \quad g\left(p_n, p_{\{l\}\neq n}\right)-g\left(p'_n, p'_{\{l\}\neq n}\right)\right),
  \end{align}
 where we can consider the prime quantities as fixed.
 
 Therefore, we can rewrite the integral \eqref{B_inv_int_2} as:
  \begin{align} \label{B_inv_int_3}
       (\sqrt{2 \pi})^{n}\int_{\mathbb{R}^n}& d^n p (1+ \beta p_r p_s \delta^{rs})^{-\frac{n}{2}-1}\psi(p) \nonumber \\
       &\times \prod_{k=1}^{n} \delta\left(V_{k}(p)\right).
   \end{align}
 where $V_{k}(p)$ are the components of the vector field $V(p)$.

 By applying the coarea formula \eqref{B_coarea}, the previous integral becomes:
  \begin{align}
      &\int_{\mathbb{R}^n} \left(\int_{V(p)=t} \!\!\!\!\!\!dH_{0}(p) \frac{(1+ \beta p_r p_s \delta^{rs})^{-\frac{n}{2}-1}\psi(p) }{\abs{\mathcal{J}_n \left[V(p)\right]}} \right. \nonumber \\
      &\left. \qquad \quad \times \prod_{k=1}^{n} \delta\left(V_{k}(p)\right) \right) d^n t   \nonumber\\
      &= \int_{\mathbb{R}^n} \!\!\,  \prod_{k=1}^n \delta(t_k) \! \left(\int_{V(p)=t} \!\!\!\!\!\!\!\!\!\!\!\!\!\! dH_{0}(p) \frac{(1+ \beta p_i p_j \delta^{ij})^{-\frac{n}{2}-1}\psi(p) }{\abs{\mathcal{J}_n \left[V(p)\right]}}\right) d^n t.
  \end{align}
 We notice that the Jacobian is well-defined and it does not have any singular points in any dimension, as can be verified by a direct calculation.
 
 The integral inside the parenthesis can be considered as function $F(t)$ to be evaluated in $t=0$ because of the presence of the deltas:
  \begin{align} \label{B_inv_int_4}
      &\int_{V(p)=0} \!\!\!\!\! dH_{0}(p) \frac{(1+ \beta p_r p_s \delta^{rs})^{-\frac{n}{2}-1}\psi(p) }{\abs{\mathcal{J}_n \left[V(p)\right]}} \nonumber \\
      & = \int_{G(p)=G(p')} \!\!\!\!\! dH_{0}(p) \frac{(1 \!+\! \beta p_r p_s \delta^{rs})^{-\frac{n}{2}-1}\psi(p) }{\abs{\mathcal{J}_n \left[V(p)\right]}}
  \end{align}

 The zero-dimensional Hausdorff measure is a discrete measure which prescribes the evaluation of the integral on the points of the domain. Since $G(p)=G(p')$ if and only if $p=p'$, the integral \eqref{B_inv_int_4} in the end is just the value of the integrand in $p=p'$.

 Therefore, coming back to the relation \eqref{B_inv_int_1}, we can write:
  \begin{align}
      &\int_{\mathbb{R}^n} \!\!\!\!\!\! d^{n} \xi \! \prod_{k=1}^{n} \! e^{-i \xi_{k} g\left(p'_k, p'_{\{l\}\neq k}\right)} \psi(\xi)\!= \!\frac{(1\!+\! \beta p'_r p'_s \delta^{rs})^{-\frac{n}{2}-1}\psi(p') }{\abs{\mathcal{J}_n \left[V(p)\right]}|_{p=p'}}
  \end{align}
 and from this we can obtain the final expression of the inverse transform, which maps wave function from quasi-position space to momentum space:
  \begin{align}
      \psi(p)= &(1+ \beta p_r p_s \delta^{rs})^{\frac{n}{2}+1} \abs{\mathcal{J}_n \left[V(p)\right]}  \nonumber \\ 
      & \times \int_{\mathbb{R}^n} d^{n} \xi \prod_{k=1}^{n} e^{-i \xi_{k} g\left(p_k, p_{\{l\}\neq k}\right)} \psi(\xi)
  \end{align}
 where we have dropped the prime.

  \section*{Acknowledgements}
     S.S. would like to thank Matteo Bruno for useful mathematical insights and discussions.

\bibliographystyle{sn-aps.bst}
\bibliography{NC_GUP}
    
\end{document}